\documentclass{IEEEtran}
\usepackage[cmex10]{amsmath}
\usepackage{url}
\usepackage{amsfonts}
\usepackage{graphicx}
\usepackage{algorithmic}
\usepackage[tight,footnotesize]{subfigure}

\newtheorem{theorem}{Theorem}[section]
\newtheorem{definition}[theorem]{Definition}
\newtheorem{proposition}[theorem]{Proposition}

\newcommand{\G}[1]{{\mathbb{#1}}}
\newcommand{\C}[1]{{\mathcal{#1}}}
\DeclareMathOperator{\adj}{adj}
\DeclareMathOperator{\cof}{cof}
\DeclareMathOperator{\col}{col}
\DeclareMathOperator{\row}{row}
\DeclareMathOperator{\Div}{div}
\newcommand{\F}{{\mathbb{F}}}
\newcommand{\Q}{{\mathcal{Q}}}
\newcommand{\N}{\mathbb{N}}

\def\udots{\mathinner{\mskip1mu\raise1pt\vbox{\kern7pt\hbox{.}}
           \mskip2mu\raise4pt\hbox{.}\mskip2mu\raise7pt\hbox{.}\mskip1mu}}

\title{Revisiting LFSRs for cryptographic applications}

\author{Fran\c cois~Arnault, Thierry~Berger, Marine~Minier and Benjamin~Pousse%
\thanks{This work was partially supported by the French National Agency of
Research: ANR-06-SETI-013.}
\thanks{Fran\c cois~Arnault, Thierry~Berger and Benjamin~Pousse are with XLIM
(UMR~CNRS~6172), Universit\'e de Limoges, 123 avenue A. Thomas, 87060 Limoges
Cedex, France, \textsf{firstname.name@xlim.fr}}%
\thanks{Marine~Minier is with  Universit\'e de Lyon, INRIA - 
INSA-Lyon, CITI, F-69621, Villeurbanne, France,
\textsf{marine.minier@insa-lyon.fr}}}

\begin{document}

\maketitle

%%%%%%% \input{abstract.tex}

%%%%%%%%%%%%%%%%%%%%%%%%%%%%%%%%%%%%%%%%
\begin{abstract}
Linear Finite State Machines (LFSMs) are particular primitives widely used in
information theory, coding theory and cryptography.  Among those linear
automata, a particular case of study is Linear Feedback Shift Registers (LFSRs)
used in many cryptographic applications such as design of stream ciphers or
pseudo-random generation. LFSRs could be seen as particular LFSMs without
inputs.

In this paper, we first recall the description of LFSMs using
traditional matrices representation.  Then, we introduce a new matrices
representation with polynomial fractional coefficients.  This new representation
leads to sparse representations and implementations.  As direct applications, we
focus our work on the Windmill LFSRs case, used for example in the E0 stream
cipher and on other general applications that use this new representation.

In a second part, a new design criterion called
diffusion delay for LFSRs is introduced and well compared with existing related notions. 
This criterion represents the diffusion capacity of an LFSR. Thus, using the 
matrices representation, we present a new algorithm to randomly pick LFSRs with
good properties (including the new one) and sparse descriptions dedicated to
hardware and software designs.  We present some examples of LFSRs generated
using our algorithm to show the relevance of our approach.

\end{abstract}

%%%%%%%%%%%%%%%%%%%%%%%%%%%%%%%%%%%%%%%%
\section{Introduction}

Linear Finite State Machines (LFSMs) are a building block of many information
theory based applications such as synchronization codes, masking or scrambling
codes. They are also used for white noise signals in communication systems,
signal sets in CDMA (Code Division Multiple Access) communications, key stream
generators in stream cipher cryptosystems, random number generators in many
cryptographic primitive algorithms, and as testing vectors in hardware design.

A Linear Finite State Machine is a linear automaton composed of memories
defined over a particular finite set $\cal{A}$ (typically a finite field) and
where the only operation updating cells is the addition
\cite{Golomb,Roggeman89,Kagaris06}. At each clock, it inputs $n$ elements of
$\cal{A}$ and outputs at least one element computed using its current state and
a linear updating function based on additions. Two main classes of LFSMs could
be defined: autonomous (without inputs in the updating process) and
non-autonomous.  
This paper first recalls the traditional representation using transition
matrices which is classically used to characterize autonomous and non-autonomous
LFSMs. Then, it introduces a new fractional representation using rational powers
series, i.e. the series are the quotient of two polynomials. Our new model
is called Rational Linear Finite State Machines (RLFSMs) and is a generalization
of the previous matrices representations. We present the link between the
two approaches. As a particular case of study of our new representation, we
focus on windmill LFSRs defined by Smeets and Chambers in \cite{Smeets88}. Those
LFSRs are based upon particular polynomials producing in parallel $v$
subsequences of a given LFSR sequence. Four windmill generators are used as
parallel updating functions in the stream cipher E0 \cite{E0}. The windmill
constructions have been first extended in \cite{Lau09a}. In this paper, we show
how we could, using the new rational representation, give a simple expression of those 
particular constructions and how this new theoretical representation could lead to clearly simplify   
the usual representation of circuits with multiple outputs at each iteration or parallelized versions of LFSRs.

In a second step, we also introduce a new criterion for LFSMs to measure what we call diffusion delay. 
We compare this new criterion with the existing notions
of auto-, cross- and simple correlations and show how this criterion captures an intrinsic behavior of the automaton itself. 
LFSMs are popular automata in many cryptographic applications and are
particularly used as updating functions of stream ciphers and of pseudo-random
generators. Their large popularity is due to their very simple design efficient
both in hardware and in software and to the proved properties of the generated
sequence (statistical properties, good periods,...) if the associated
polynomial is primitive. In many cryptographic applications, the diffusion delay of
  LFSMs is most of the time not considered. In this paper, we focus on
this criterion, show its link with correlation and its effectiveness for several types of automata such as FCSRs or NLFSRs. We also give a new algorithm to construct hardware and/or software
efficient LFSMs with good diffusion delay called Ring LFSRs. For the hardware case, we
show theoretical bounds on the number of gates required to implement a ring
LFSR compared with the traditional Galois and Fibonacci LFSRs and we compare the
associated traditional properties. For the software case, we compare the
properties and the performances of our Ring LFSR with the LFSR involved in the
stream ciphers SNOW v2.0 \cite{EJ02}, finalist of the NESSIE project
\cite{Ness01a}.

This paper is organized as follows: Section \ref{sec-background} gives some
background about Finite State Machines (FSMs) and introduces notations. Section
\ref{sec-previous} presents previous works on LFSMs. Section \ref{sec-poly}
introduces the new rational representation for LFSMs, detailing some examples of Windmill LFSRs and of general applications. Section \ref{sec-implem} presents the new diffusion delay criterion, shows why this criterion captures new notions and proposes hardware and software oriented implementations with respect to
this criterion. Finally, Section \ref{sec-conclu} concludes this paper.

\subsection{Notations}

  The finite field with cardinal~$q$ is denoted~$\F_q$. 
We denote $\F_q[X]$ the ring of polynomials and $\F_q[[X]]$ the ring of power
series, both over
$\F_q$.  We will also use in Sections~\ref{sec-poly} and followings, the
ring~$\Q$ of rational power 
series, that is the ring of power series which can be written $P(X)/Q(X)$ where
$P,Q\in\F_q[X]$ with 
$Q(0)\neq0$.  We will recall in Theorem~\ref{theo-ratpowerseries} that $\Q$ is
the ring of power 
series that correspond to eventually periodic sequences.

  We will also use the notation $\C M_{k,l}(\C R)$ for the ring of matrices with
$k$ rows and $l$ columns over a ring $\C R$.   For convenience and not to make
notations too heavy,
we often write vectors $v$ as rows $v=(v_1,\ldots,v_n)$ but also use them as
column vectors
in expressions such as $Av$ where $A$ is a matrix.  Of course the correct form
should be with
explicit transposition as in $A\,{}^tv$ but we expect the reader not to be
confused with this abuse
of notation.

  In Section~\ref{sec-implem}, we will use the notation $w_H$ for the Hamming
weight.
For example, the Hamming weight of a matrix is its number of nonzero entries.
The Hamming weight
of a polynomial is its number of non null coefficients.

\section{Background}\label{sec-background}

\subsection{Linear recurring sequences}

As the case of binary sequences is the most useful in pseudo-random generation,
we deal in this
paper with the two elements field~$\F_2$.  However most of the results presented
here have a
straightforward generalization when using another finite field as base field.

  Recall that a sequence $s=(s_i)_{i\in\N}$ over $\F_2$ is a {\it linear
recurring sequence} if
there exists $q_1,\ldots,q_d\in\F_2$ such that $s_n=q_1 s_{n-1}+\cdots+q_d
s_{n-d}$ for all 
$n\geq d$.   A binary sequence $(s_i)_{i\in\N}$ can be seen as a power series
$s(X)=\sum_{i=0}^\infty
s_iX^i$.  In terms of power series, we have the following Theorem \cite{Golomb}:

\begin{theorem}
\label{theo-ratpowerseries}
Let $s=(s_i)_{i\in\N}$ be a sequence over $\F_2$. The following statements are
equivalent:
\begin{itemize}
\item The sequence $s$ is a linear recurring sequence.
\item The sequence $s$ is eventually periodic,  i.e.~there exists $N\in\N$ such
that $(s_i)_{i\ge
N}$ is periodic.
\item There exist polynomials $f(X),g(X)\in\F_2[X]$ with $g(0)=1$ such that the
power series
  $f(X)/g(X)$ is equal to $\sum_{i\in\N}s_iX^i$, i.e. $s(X)$ is in $\Q$.
\end{itemize}
Moreover, $s$ is periodic if and only if $f(X)$ and $g(X)$ are such that $\deg
f<\deg g$.
\end{theorem}

  According to this Theorem a correspondence can be built between rational
power series and
sequences.  The period of a linear recurring sequence is determined by the
polynomial~$g(X)$ as
shown by the following Theorem~\cite{Golomb}:
\begin{theorem}\label{theo-mseq}
Let $s(X)=f(X)/g(X)$ be a rational power series, with $\gcd(f(x),g(x))=1$.  We
denote by $s$ the
sequence of coefficients of $s(X)$.

\begin{itemize}
\item The period of $s$ is equal to the order of $X$ in $\F_2[X]/(g(X))$.
\item If $g(X)$ is primitive then there exists $N\in\N$ such that $\sum_{i\ge
N}s_iX^{i-N}=1/g(X)$.
\end{itemize}

\end{theorem}

When the polynomial $g(X)$ is primitive, the sequence $s$ has period $2^{\deg
g}-1$ and is called a
$m$-sequence.

\subsection{Adjunct matrix}

  Let $M=(m_{i,j})_{1\leq i,j\leq n} $ be a square matrix over a ring~$\C R$.
The $(i,j)$-th cofactor
$c_{i,j}$ of $M$ is $(-1)^{i+j}$ times the determinant of the matrix obtained
by removing the line~$i$
and the column~$j$ in~$M$.  The transpose of the cofactor matrix~$(c_{i,j})$ is
called the 
{\it adjunct matrix} of~$M$ and we denote it by~$\adj(M)$.  The adjunct
of~$M$ has its
coefficients in~$\C R$ and satisfies the following identity
\begin{equation}
  \adj(M) M = M\adj(M) = \det(M) I.
  \label{adjugate}
\end{equation}
Hence, if $\det(M)$ is invertible, we have $M^{-1}=\frac1{\det(M)}\adj(M)$.

\section{LFSMs}
\label{sec-previous}

\subsection{Definitions}

LFSMs (Linear Feedback State Machines) have been studied in
\cite{Stone73,Golomb,Roggeman89,Klapper}.   They are a generalization of Linear
Feedback Shift
Registers, for which the shift structure is removed, i.e. each cell has no
privileged
neighbor.  Let us give a definition of an LFSM (over~$\F_2$):

\begin{definition}\label{def-lfsm}
A Linear Finite State Machine (LFSM) $\C L$, of length~$n$, with $k$ inputs and
$\ell$
outputs consists of:
\begin{itemize}
\item A set of $n$ cells, each of them storing a value in~$\F_2$.  The content
of the cells, a
binary vector of length~$n$, will be denoted $m=(m_0,\ldots,m_{n-1})$ and is
called the {\it state}
of the LFSM.  We will sometimes call the set of these $n$ cells the {\it
register}.
\item A {\it transition function} which is a linear function from
$\F_2^n\times\F_2^k$ to $\F_2^n$.
\item An {\it extraction function} which is a linear function from $\F_2^n$
to~$\F_2^\ell$.
\end{itemize}
\end{definition}

\noindent   The behavior of an LFSM is described below:
\begin{enumerate}
\item[1] The register is initialized to a state $m^{(0)}\in\F_2^n$ at time
$t\leftarrow0$.
\item[2] The extraction function is used to compute an output vector
$v(t)\in\F_2^\ell$ from the
state~$m^{(t)}$.
\item[3] A new state $m^{(t+1)}$ is computed from the current state~$m^{(t)}$
and from a vector~$u^{(t)}\in\F_2^k$ input at time~$t$ using the transition
function.
This new state is stored in the register.
\item[4] Execution continues by going back to Step 2, with $t\leftarrow t+1$.
\end{enumerate}

An LFSM is a kind of finite state automaton, for which the set of states is
$\F_2^n$ and the
transition function is linear.  However, an additional function gives the
ability to output data.
An LFSM is also different from a finite state automaton because the transition
function may depend
also of an input vector.  Note also that an LFSM does not terminate as it has no
final state.

  A given LFSM can be entirely specified by a triplet of $\F_2$-matrices
$(A,B,C)$, of respective
sizes $n\times n$, $n\times k$ and $\ell\times n$, which describe the
transition and extraction
functions in the following way.  Given a state column vector $m^{(t)}\in\F_2^n$
and an
input column vector $u^{(t)}\in\F_2^k$, the next state vector $m^{(t+1)}$ and
the present output
vector $v^{(t)}\in\F_2^{\ell}$ are expressed by:  
\begin{eqnarray}
  m^{(t+1)} &=& Am^{(t)}+Bu^{(t)},
  \label{transition}\\
  v^{(t)}   &=& Cm^{(t)}.
  \label{extraction}
\end{eqnarray}
For suitable matrices $A,B,C$, we will denote $\C L(A,B,C)$ an LFSM with
transition and extraction
functions given by Equations \ref{transition} and~\ref{extraction}.
For short, we will often call $A$ the transition matrix of $\C L$ (even
when $B\neq0$) while in fact the transition function depends on both $A$
and~$B$.

The polynomial defined now plays an important role in the theory of LFSMs:
\begin{definition}
\label{def-feedpoly1}
Let $\C L=(A,B,C)$ be an LFSM.  The polynomial $\det(I-XA)$ is called the {\it
connection
polynomial} of $\C L$.  We will denoted it $Q_{\C L}(X)$ or simply~$Q(X)$.
\end{definition}

  Note that $Q(X)\in\F_2[X]$ has degree at most~$n$ (with equality iff $\det
(A)\neq0$).  Moreover,
$Q(0)=1$, hence $Q(X)$ has an inverse in the ring $\F_2[[X]]$ of power
series. More precisely, $Q(X)^{-1}$ is in $\Q$.

\subsection{Sequences obtained from an LFSM}

  For each $t\in\N$, an LFSM outputs a vector
$v^{(t)}=(v_1^{(t)},\ldots,v_\ell^{(t)})$ of $\ell$
bits.  For each $i=1,\ldots,\ell$, we will denote $V_i(t_0)=\sum_t^\infty
v_i^{(t_0+t)}X^t$ the power
series obtained from the sequence $(v_i^{(t)})_{t\geq t_0}$.  We also define
$V^{(t_0)}$ as the vector
$(V_1(t_0),\ldots,V_\ell(t_0))$ of power series.  We consider also the series 
$M_i(t_0)=\sum_t^\infty m_i^{(t_0+t)}X^t$ obtained from the sequence observed
in each cell $m_i$
(for $1\leq i\leq n$), and the vector $M^{(t_0)}=(M_1(t_0),\ldots,M_n(t_0))$ of
power series.  In a
similar way, we define $U^{(t_0)}=(U_1(t_0),\ldots,U_k(t_0))$ from the input
sequences.

  The sequences $M_i(t_0)$ observed in the register, and the output sequences
$V_i(t_0)$ satisfy
interesting linear relations (cf.~\cite{Golomb,Stone73,Kagaris06}).   We
provide these relations in the next
theorem.

\begin{theorem}\label{theo-lfsm}
Let $\C L=(A,B,C)$ be an LFSM. The vectors $M^{(t_0)}$ and $V^{(t_0)}$ verify:
\[
  \left\{
  \begin{array}{l}
     M^{(t_0)}
     =
     \displaystyle\frac{\adj(I-XA)}{Q_{\C L}(X)} (m^{(t_0)}+XBU^{(t_0)})\\
     V^{(t_0)}
     =
     \displaystyle C \frac{\adj(I-XA)}{Q_{\C L}(X)} (m^{(t_0)}+XBU^{(t_0)}).
  \end{array}
  \right.
\]
\end{theorem}

\begin{IEEEproof}
For each $t\in\N$, we multiply Equation~\ref{transition} and
Equation~\ref{extraction} by $X^t$
and sum each of them over $t$.  We get
\begin{eqnarray}
  M^{(t_0+1)}  &=&   AM^{(t_0)}+BU^{(t_0)}
  \label{eqn1-theo-lfsm} \\
  V^{(t_0)}    &=&   CM^{(t_0)}.
  \label{eqnx-theo-lfsm}
\end{eqnarray}
But $M^{(t_0)}=m^{(t_0)}+XM^{(t_0+1)}$. 
Hence, with Equation~\ref{eqn1-theo-lfsm} we obtain
$$
  M^{(t_0)} = X(AM^{(t_0)} + BU^{(t_0)}) + m^{(t_0)}
$$
or also
$
  (I - XA) M^{(t_0)} = XBU^{(t_0)} + m^{(t_0)}
$.
By Equation~\ref{adjugate} we obtain the first relation of
Theorem~\ref{theo-lfsm}.
The second one follows from Equation~\ref{eqnx-theo-lfsm}. 
\end{IEEEproof}

Note that, as mentioned before, $1/Q_{\C L}(X)$ is a power series.  So the
expression given for
$M^{(t_0)}$ in Theorem~\ref{theo-lfsm} does not (in general) belong
to~$\F_2[X]$ but to
$\F_2[[X]]$, even if
the input $U$ is of finite degree.

  Note also that, when the LFSM $\C L$ has no input (or more generally when the
input $U$ has finite
degree), Theorem~\ref{theo-lfsm} gives expressions for $M_i^{(t_0)}$ and
$V_i^{(t_0)}$ as quotients
of two polynomials, and so belong to $\Q$, the ring of rational power series.

\subsection{Autonomous LFSMs}

  An important particular case of LFSMs is the one for which the transition
function does not depend
on some input, that is to say~$B=0$.  Such an LFSM will be called an autonomous
LFSM.  
The following Theorem shows that some polynomials $p_i$ (for $1\leq i\leq n$)
related to the
components $m_i$ of the state are divided by~$X$ modulo~$Q(X)$ at each clock
cycle.

\begin{theorem}
Let $\C L$ be an autonomous LFSM and put $p^{(t)}=\adj(I-XA)m^{(t)}$ (for
$t\in\N$).  The relation
$Xp^{(t+1)}\equiv p^{(t)}$ modulo~$Q(X)$ holds, for each $t$.
\end{theorem}

\begin{IEEEproof}
From Equation~\ref{transition}, we have
$
  Xm^{(t+1)} 
  = XAm^{(t)}
  = -(I - XA)m^{(t)} + m^{(t)}
$.
Multiplication by~$\adj(I-XA)$ gives $Xp^{(t+1)}=-Q(X)m^{(t)}+p^{(t)}$.
\end{IEEEproof}

\subsection{Similar LFSMs}

  Two LFSMs defined by two distinct triples $(A,B,C)$ and $(A',B',C')$ may
produce the same output.
This is the case of {\it similar} LFSMs, which were defined
in~\cite{Kagaris06,Stone73}.

\begin{definition}
\label{def-lfsm-similar}
Given two LFSMs $\C L=(A,B,C)$ and $\C L'=(A',B',C')$. $\C L$ and $\C L'$ are
said similar if there exists a non-singular matrix $P$ over $\F_2$ such that:
\[
  A'=P^{-1}AP,
  \qquad
  B'=P^{-1}B,
  \qquad
  C'=CP.
\]
The matrix $P$ is called the {\it change basis matrix from $\C L$ to $\C L'$}.
\end{definition}

\begin{theorem}
\label{theo-lfsm-similar}
Let $\C L$ and $\C L'$ be two similar LFSMs. Assume that their initial state
vectors satisfy
$m'^{(0)}=P^{-1}m^{(0)}$ and that they have same input ($U^{(0)}=U'^{(0)}$).
Then:
\begin{enumerate}
\item Both LFSMs $\C L$ and $\C L'$ have same connection polynomial.
\item $M'^{(0)}=P^{-1}M^{(0)}$. In particular, $m'^{(t)}=P^{-1}m^{(t)}$ holds
for each $t\ge 0$.
\item The sequences output by $\C L$ and $\C L'$ are equal:
$V'^{(0)}=V^{(0)}$. In particular, $v'^{(t)}=v^{(t)}$ holds for each $t\ge 0$.
\end{enumerate}
\end{theorem}

\begin{IEEEproof}
\begin{enumerate}

\item The first claim results from 
$
  \det(I-XA') = \det(I-XP^{-1}AP) = \det(P^{-1}(I-XA)P) = \det(I-XA).
$

\item Let's prove the second claim by recurrence.  If $m'^{(t)}=P^{-1}m^{(t)}$
for some $t$, then Equation~\ref{transition} gives
$ P^{-1}m^{(t+1)}
  =
  P^{-1}Am^{(t)} + P^{-1}Bu^{(t)}
  =
  P^{-1}AP m'^{(t)} + P^{-1}Bu^{(t)}
  =
  A'm'^{(t)} + B'u'^{(t)}
  =
  m'^{(t+1)}
$. 

\item Finally, using Equation~\ref{extraction},
$
  v'^{(t)}
  =
  C'm'^{(t)}
  =
  CPP^{-1}m^{(t)}
  =
  Cm^{(t)}
  =
  v^{(t)}
$. This proves the last claim.
\end{enumerate}

\end{IEEEproof}

\subsection{Classical families of autonomous LFSMs}

Different special cases of LFSMs, are well-known for years and have been
extensively studied, with
some variations of terminology among different scientific communities, for
example the theoretic
and electronic communities as~\cite{Stone73,Kagaris06,Cattell98} and the
cryptographic community
as~\cite{Goldberg96,Leglise05,Joux06,Klapper}.  We gather in this subsection
some of these special
cases, using notations consistent with the one we used above. 

\begin{figure}[!t]
\centering
\subfigure[Galois LFSR]{
\includegraphics[width=0.95\columnwidth]{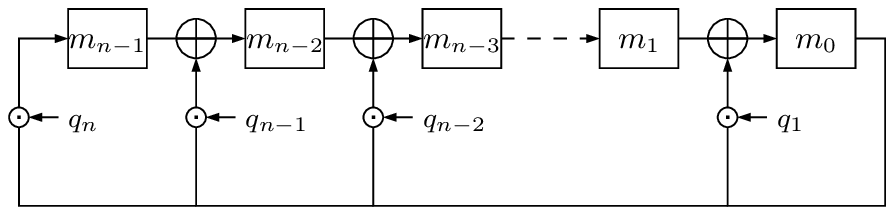}
}
\subfigure[Fibonacci LFSR]{
\includegraphics[width=0.95\columnwidth]{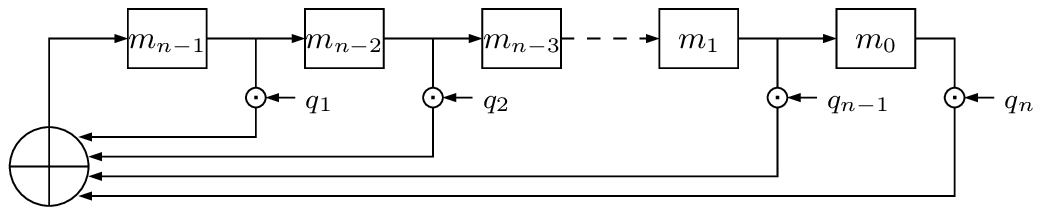}
}
\caption{Implementation of Galois and Fibonacci LFSRs with connection polynomial
$Q(X)=q_nX^n+\cdots+q_1X+1$}\label{fig-galfib}
\end{figure}

The most famous LFSMs special cases are:
\begin{itemize}
\item the \emph{Fibonacci Linear Feedback Shift Registers}, also known as
\emph{External-XOR LFSR}, or just \emph{LFSR};
\item the \emph{Galois Linear Feedback Shift Registers}, also known as
\emph{Internal-XOR LFSR}, or \emph{Canonical LFSR}.
\end{itemize}
A Galois or Fibonacci LFSR is defined by its connection polynomial because the
transition matrix~$A$
has a special form and can be deduced from it.  The matrices $B$ and~$C$ are
simple because LFSR
have no input and because they output a single bit.  The transition matrices
for Galois and
Fibonacci are shown in Figure~\ref{fig-trans}. Figure~\ref{fig-galfib} presents
the corresponding
implementations.  

  It can be shown that the matrices $T_F$ and $T_G$ given in
Figure~\ref{fig-trans} are {\it
similar} matrices (because they are ``transposed with respect to the second
diagonal'' one from each other).
Hence, the Galois and Fibonacci LFSRs with same connection polynomial are
similar
LFSMs in the sense of Definition~\ref{def-lfsm-similar}.

\begin{figure}[!t]
\centering
\subfigure[Galois LFSR]{
$T_G=\begin{pmatrix}
q_1&	1&		&		&		\\
q_2&	&		1&		(0)&	\\
\vdots&	&		(0)&	\ddots&	\\
q_{n-1}&&		&		&	1	\\
q_n&	0&		0&		\cdots&	0
\end{pmatrix}$
}
\subfigure[Fibonacci LFSR]{
$T_F=\begin{pmatrix}
0&		1&		&		&		\\
0&		&		1&		(0)&	\\
\vdots&	&		(0)&	\ddots&	\\
0&		&		&		&		1\\
q_n&	q_{n-1}&\cdots&	q_2&	q_1
\end{pmatrix}$}
\caption{Transition matrices of Galois and Fibonacci LFSRs with connection
polynomial $Q(X)=q_nX^n+\cdots+q_1X+1$}\label{fig-trans}
\end{figure}

\begin{figure}[!t]
\centering
\subfigure[Transition matrix of a CA]{
$T_{CA}=\begin{pmatrix}
q_1&	1&		&		&		\\
1&		q_2&	1&		(0)&	\\
&		\ddots&	\ddots&	\ddots&	\\
&		(0)&	1&		q_{n-1}&	1\\
&		&		&		1&		q_n
\end{pmatrix}$
}
\subfigure[Implementation of a CA]{
\includegraphics[width=0.95\columnwidth]{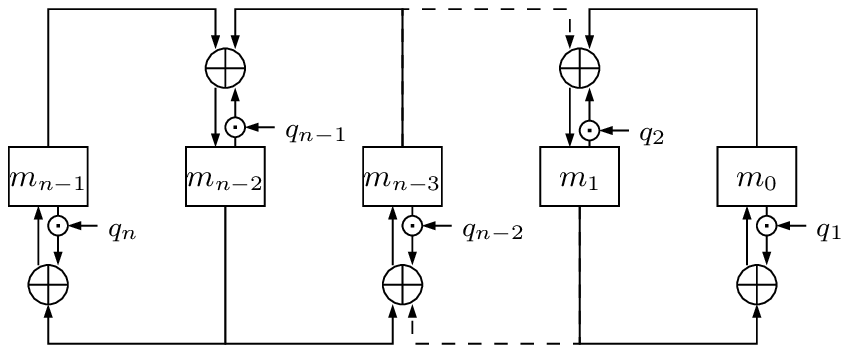}
}
\caption{Transition matrix and implementation of a 3-neighborhood Cellular
Automaton}\label{fig-CA}
\end{figure}

Another special kind of LFSMs is the 3-neighborhood cellular automaton (CA)
\cite{Cattell98,Cattell96a,Cattell96b,Kagaris06}. These automata are
characterized by a
tri-diagonal matrix as presented in Figure~\ref{fig-CA}. They are suitable for
hardware
implementation. 

To cover numerous kind of automata presented in
\cite{Kagaris06,Polonais,Cattell96b,Dufaza98}, we
introduce Ring LFSRs.  The cells which store the state are organized in a cyclic
shift
register. This corresponds to a transition matrix of a particular form: 
\begin{definition} \label{def-lf}
An LFSM $\C L$ with transition matrix $A$ is called a {\it Ring Linear Feedback
Shift Register} if
$A=(a_{i,j})_{0\le i,j<n}$ as the following form:
\[
\left\{\begin{array}{l}
a_{i,i+1}=1\text{ for all }0\le i<n-1\\
a_{n-1,0}=1
\end{array}\right.
\]
i.e.,
\[A=\begin{pmatrix}
      & 1    &      & (*)  &      \\
      &      &\ddots&      &      \\
      &  (*) &      &\ddots&      \\
      &      &      &      & 1    \\
    1 &      &      &      &      \\
\end{pmatrix}
\]
\end{definition}

In particular, Galois and Fibonacci LFSRs are special cases of Ring LFSRs.

\begin{figure}[!t]
\centering
\subfigure[Galois LFSR $\C
L_0$]{\includegraphics[width=0.95\columnwidth]{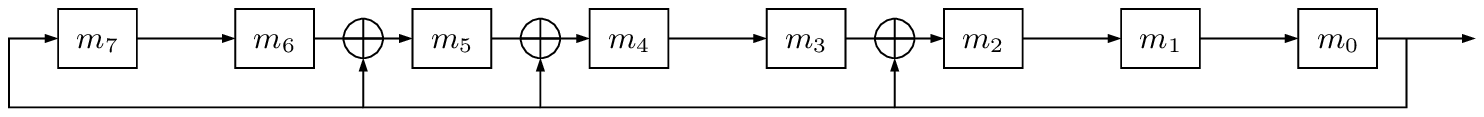}}
\subfigure[Fibonacci LFSR $\C
L_1$]{\includegraphics[width=0.95\columnwidth]{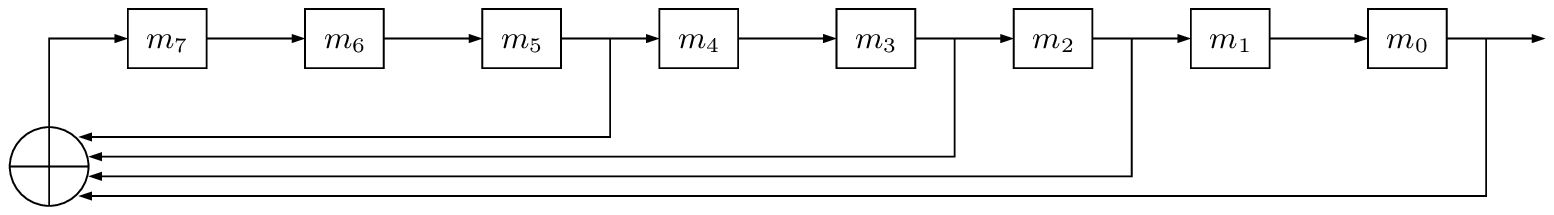}}
\subfigure[Ring LFSR $\C
L_2$]{\includegraphics[width=0.95\columnwidth]{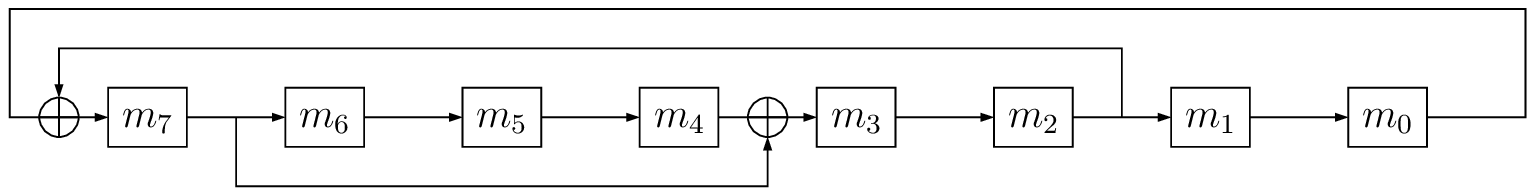}}
\caption{Three LFSR with connection polynomial
$Q(X)=X^8+X^6+X^5+X^3+1$}\label{fig-hard-ex}
\end{figure}

We detail here a complete example of these automata. Consider the primitive
connection polynomial
$Q(X)=X^8+X^6+X^5+X^3+1$.  Denote $\C L_0$ the associated Galois LFSR, $\C L_1$
the associated
Fibonacci LFSR and $\C L_2$ a generic Ring LFSR with connection
polynomial~$Q(X)$.  We present their
respective transition matrices $T_0$, $T_1$ and $T_2$ in
Figure~\ref{fig-trans-ex}.
Figure~\ref{fig-hard-ex} shows the implementation of $\C L_0$, $\C L_1$ and $\C
L_2$ whereas
Table~\ref{tab-hard-ex} displays the states of these automata during 8 clocks
starting from the
same initial state.

\begin{figure}[!t]
\centering
\subfigure[Galois LFSR]{
$T_0=\begin{pmatrix}
0&1& & & & & & \\
0& &1& & &(0)& & \\
1& & &1& & & & \\
0& & & &1& & &\\
1& & & &&1& & \\
1& &(0)& & & &1& \\
0& & & & & & &1\\
1& & & & & & & \\
\end{pmatrix}$
}
\subfigure[Fibonacci LFSR]{
$T_1=\begin{pmatrix}
 &1& & & & & & \\
 & &1& & &(0)& & \\
 & & &1& & & & \\
 & & & &1& & &\\
 & & & &&1& & \\
 & &(0)& & & &1& \\
 & & & & & & &1\\
1&0&1&1&0&1&0&0\\
\end{pmatrix}$}
\subfigure[Ring LFSR]{
$T_2=\begin{pmatrix}
 &1& & & & & & \\
 & &1& & &(0)& & \\
 & & &1& & & & \\
 & & & &1& & &1\\
 & & & & &1& & \\
 & &(0)& & & &1& \\
 & & & & & & &1\\
1& &1& & & & & \\
\end{pmatrix}$
}
\caption{Transition matrices of $\C L_0$, $\C L_1$ and $\C
L_2$}\label{fig-trans-ex}
\end{figure}

\begin{table}
\[
\begin{array}{|c||c|c|c|}
\hline
	&	\C L_0	&	\C L_1	&	\C L_2\\
\cline{2-4}
Clock	&	Cells&Cells&Cells\\
	&	7654321\ 0&7654321\ 0&7654321\ 0\\
\hline
\hline
0	&	\texttt{0000000\ 1}&\texttt{0000000\ 1}&\texttt{0000000\ 1}\\
1	&	\texttt{1011010\ 0}&\texttt{1000000\ 0}&\texttt{1000000\ 0}\\
2	&	\texttt{0101101\ 0}&\texttt{0100000\ 0}&\texttt{0100100\ 0}\\
3	&	\texttt{0010110\ 1}&\texttt{0010000\ 0}&\texttt{0010010\ 0}\\
4	&	\texttt{1010001\ 0}&\texttt{1001000\ 0}&\texttt{1001001\ 0}\\
5	&	\texttt{0101000\ 1}&\texttt{0100100\ 0}&\texttt{0100000\ 1}\\
6	&	\texttt{1001110\ 0}&\texttt{1010010\ 0}&\texttt{1010000\ 0}\\
7	&	\texttt{0100111\ 0}&\texttt{0101001\ 0}&\texttt{0101100\ 0}\\
8	&	\texttt{0010011\ 1}&\texttt{0010100\ 1}&\texttt{0010110\ 0}\\
\hline
\end{array}
\]
\caption{States of $\C L_0$, $\C L_1$ and $\C L_2$ during 8
clocks.}\label{tab-hard-ex}
\end{table}

The reader can see that from the same initial state \texttt{00000001} the
output sequences are distinct.  However, they are all 
a part of the same
$m$-sequence defined by $Q(X)=X^8+X^6+X^5+X^3+1$ according to
Theorem~\ref{theo-lfsm}.  In other
words there exists three different polynomials $P_0(X),P_1(X),P_2(X)$ of degrees
less than~8 such
that the sequences generated by $\C L_0$, $\C L_1$ and $\C L_2$ are respectively
$P_0(X)/Q(X)$,
$P_1(X)/Q(X)$ and $P_2(X)/Q(X)$.

%%%%%%%%%%%%%%%%%%%%%%%%%%%%%%%%%%%%%%%%

\section{Rational representation}\label{sec-poly}

In this section, we will introduce a generalization of LFSRs and LFSMs by 
extending the set of possible coefficients for the transition matrix to rational fractions. This new 
approach is not only of theoretical interest, but is also an interesting tool for both
having a more global view of complex circuits and for constructing more complex 
circuits from smaller LFSMs with nice properties. 
Each coefficient of such a matrix is a rational fraction which represents a small 
LFSM. The inputs and outputs of each small LFSM are thus used as a part of the full automaton.

This new representation allows an easier description of complex circuits with small internal 
components such as the so-called Windmill generators \cite{Smeets88}. These generators are for example used in the stream cipher E0 \cite{E0} 
implemented in the Bluetooth system.

This rational representation leads to a simpler representation of some circuits with 
multiple outputs at each iteration or of parallelized versions of LFSRs.

This section is organized as follows:  we first focus our analysis on LFSMs with a single input and a single output. 
Then we introduce the notion of transition matrix with rational coefficients. We demonstrate that the automata built using this new representation 
essentially produce the same sequences than the classical LFSRs. We give a first example based on this new representation to construct a filtered LFSR automaton.
We then focus our work on the case of Windmill generators and give a simpler and more compact definition of such LFSRs. We thus discuss the 
difficulty of implementing such automata which is not so easy in the general case. Finally, 
we conclude this section with a concrete example. It consists in 
a generalization of Windmill generators that allows to construct complex 
circuits from simpler well designed circuits. These simple circuits are building 
blocks of a bigger automaton which connects the small components in a circular way. The full circuit inherits good 
internal properties of the smaller components.

%%%%%%%%%%%%%%%%%%%%%%%%%%%%%%%%%%%%%%%%

\subsection{LFSMs with a single input and a single output}\label{1to1}

As a building block for our representation, we are first interested by an LFSM
with
a single input bit and a single output bit. In  this situation, the matrix $B$ 
is a $n\times 1$ matrix, with a single 1 in position $i_0$. Likewise, $C$ is 
a $1\times n$ matrix, with a single 1 in position $j_0$.

Set $A'=\adj(I-XA)=(A'_{i,j}(X))$, where the coefficients $A'_{i,j}(X)$ are 
polynomials, and $Q(X)=\det(I-X.A)$.  We can derive
from 
Theorem \ref{theo-lfsm}, the following relation 
between the input series $U^{(t)}$ and the output series $V^{(t)}$:

$$ V^{(t)}= \frac{X}{Q(X)}CA'BU^{(t)}+\frac{1}{Q(X)}CA'm^{(t)}$$

Note that $CA'B=A'_{j_0,i_0}(X)$ is a polynomial, and $P^{(t)}(X)=CA'm^{(t)}$
is 
also a polynomial. Setting $R(X)=XA'_{i_0,j_0}(X)$, we can rewrite the previous 
formula
$$ V^{(t)}= \frac{R(X)}{Q(X)}U^{(t)}+\frac{P^{(t)}(X)}{Q(X)}$$

Note that $R(X)$ is independent of the internal state $m^{(t)}$ of the LFSM, 
and $\frac{P^{(t)}(X)}{Q(X)}$ is uniquely determined by the internal state
$m^{(t)}$
of the LFSM.

So up to initial internal values of such LFSM, we can consider that it performs 
the multiplication of the input by the rational series $R(X)/Q(X)$
(note that, since $Q(X)=\det(I-X.A)$, we have $Q(0)=1\neq 0$).

Conversely, for a given rational power series $R(X)/Q(X)$, $Q(0)\neq 0$, it is 
possible to construct many LFSMs which perform the multiplication by
$R(X)/Q(X)$.

\begin{figure}[!t]
\centering
\includegraphics[width=0.95\columnwidth]{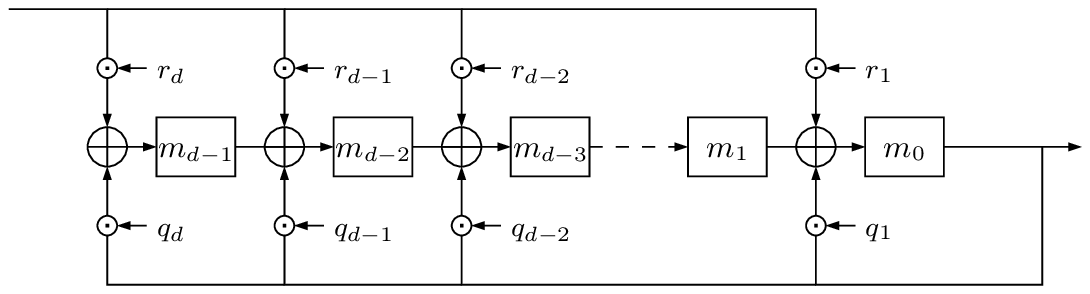}
\caption{Implementation of a division/multiplication circuit}\label{fig-vane}
\end{figure}

As an example of such LFSMs, we give in Figure~\ref{fig-vane} an LFSM with one 
input and one output which performs the multiplication by $R(X)/Q(X)$ called in
the rest of this paper a Galois vane (in reference to a Galois LFSR and a vane
of a windmill generator). 

The matrix description of this LFSM is:
\[
A=\begin{pmatrix}
q_1&	1&		&		&		\\
q_2&	&		1&		(0)&	\\
\vdots&	&		(0)&	\ddots&	\\
q_{d-1}&&		&		&	1	\\
q_d&	0&		0&		\cdots&	0
\end{pmatrix},
B=\begin{pmatrix}
r_1\\
r_2\\
\vdots\\
r_{d-1}\\
r_d
\end{pmatrix}
\]
and $C=(1,0,\ldots,0)$.

it will be interesting to use some multiplication/division circuits  which are
not performed by a Galois vane. As an example,  we consider the 
ring LFSR described in Figure \ref{fig-hard-ex}. The connection polynomial is
$Q(X)=X^8+X^6+X^5+X^3+1$. Let $T'=\adj{I-XT_2}$, we have $T'_{1,1}=X^6+X^3+1$
and $T'_{4,3}=X^7+X^5+X^4+X^2$. For ${}^tB=C=(1,0,...,0)$, this ring LFSR
performs the multiplication by $(X^6+X^3+1)/(X^8+X^6+X^5+X^3+1)$. For
${}^tB=(0,0,0,1,0,0,0,0)$ and $C=(0,0,1,0,...,0)$, it performs the
multiplication by $(X^7+X^5+X^4+X^2)/(X^8+X^6+X^5+X^3+1)$. For these two
examples, the circuit is simpler than the equivalent one obtained by the
Galois vane.

\subsection{Rational Linear Machines}

Now, we want to use multiplications by rational power series $R(X)/Q(X)$, 
with $Q(0)\neq0$, as internal building blocks in order to construct bigger
LFSMs. 

Recall that we denote by $\Q$ the ring of rational power series, that is 
$\{P(X)/Q(X)\in\F_2[[X]]\mid P(X),Q(X)\in\F_2[X], Q(0)\neq0\}$.

\begin{definition}\label{def-rlm}
A Rational Linear Machine (RLM) $\C L$ with $k$-bit input, $\ell$-bit
output and length $n$ over $\Q$ is a triplet of matrices $(A,B,C)$ over
 $\Q$, of respective sizes $n\times n$, $n\times k$, $\ell\times n$.
Given the current state vector $(m^{(t)},c^{(t)})\in\C M_{n,1}( \F_2)\times\C
M_{n,1}(\C Q)$ and input vector $u^{(t)}\in\C M_{k,1}(\G F_2)$. The next
state vector $(m^{(t+1)},c^{(t+1)})$ and the present output vector
$v^{(t)}\in\C M_{\ell,1}(\G F_2)$ are expressed as:
\[
\left\{\begin{array}{rcl}
m^{(t+1)}&=&Am^{(t)}+c^{(t)}+Bu^{(t)}\bmod X\\
c^{(t+1)}&=&Am^{(t)}+c^{(t)}+Bu^{(t)}\Div X\\
v^{(t)}&=&Cm^{(t)}
\end{array}\right.
\]
where $P(X)\Div X=\frac{P(X)-(P(X)\bmod X)}{X}$.
\end{definition}

As previously we are able to describe the output sequences:

\begin{theorem} \label{RLM}
Let $\C L=(A,B,C)$ a RLM. The vector $M^{(t)}$ satisfy the relation:
$$
M^{(t)}=(I-X A)^{-1} \left(m^{(t)}+Xc^{(t)}+X B U^{(t)}\right)
$$
\end{theorem}

\begin{IEEEproof}
With the previous notations we have the following relations:
\begin{eqnarray}
M^{(t+1)}&=&A M^{(t)}+c^{(t)}+BU^{(t)}\label{eqn1-theo-lfsr2}\\
M^{(t)}&=&X M^{(t+1)}+m^{(t)}\label{eqn2-theo-lfsr2}
\end{eqnarray}

Equation~\ref{eqn1-theo-lfsr2} is by Definition~\ref{def-rlm}.
Equation~\ref{eqn2-theo-lfsr2} comes from the Definition of $M^{(t)}$. It leads
to the
following relation:
\[
(I-X A) M^{(t_0)}=m^{(t_0)}+Xc^{(t_0)}+X B U^{(t)}
\]
Note that $(I-X A)$ is invertible in $\C M_n(\Q)$. This leads to
$M^{(t_0)}=(I-X A)^{-1}(m^{(t_0)}+Xc^{(t_0)}+X B U^{(t)})$
in $\Q$.
\end{IEEEproof}

%%%%%%%%%%%%%%%%%%%%%%%%%%%%%%%%%%%%%%%%
\subsection{Rational Linear Finite State Machines}

In order to focus the attention on some applications, and for a better 
understanding of the significance of Theorem \ref{RLM}, we focus in this 
Section on the study of RLM with no input. Moreover, we will try to limit the 
domain of the ``carries'' register $c$ in order to ensure that the machine is a
finite 
state machine.
We suppose in the sequel that $B=0$, i.e. there is no input.

In order to restrict RLM to finite state machines, we have to look at the
evolution of ``internal memories'' $c^{(t)}$ in 
more details. Let 
$A_{i,j}=P_{i,j}(X)/Q_{i,j}(X)$ be the expression of a coefficient of the 
matrix $A$ as a quotient of two polynomials. For a fixed row $i$ we can compute
the polynomial $Q_i(X)=\mbox{lcm}(Q_{i,1}(X),\dots,Q_{i,n}(X))$. So we can
normalize the 
rational representations as follows:  $A_{i,j}=R_{i,j}(X)/Q_i(X)$.  
For each 
row $i$ we define the following finite subset of $\Q$:  $W_i=\{R(X)/Q_i(X)\:|\: 
\deg(R(X))<\max_j(\deg(R_{i,j}(X)))\}$. Finally we define $W=\prod_{i=1}^{n}W_i 
\subset \Q^n$. Note that $W$ is a finite set. The following proposition shows
that it is a ``reasonable'' set for the values of the internal memories;
\begin{proposition}
    Suppose that at time $t_0$, $c^{(t_0)}$ is in $W$, then for any $t\geq 
    t_{0}$, $c^{(t)}$ is in $W$.
\end{proposition}

\begin{IEEEproof}
    Let $\mu^{(t+1)}=Am^{(t)}+c^{(t)}$. From the definition of a RLM, we have 
  $m^{(t+1)}=  \mu^{(t+1)}\bmod X$ and $c^{(t+1)}=  \mu^{(t+1)}\Div X$.
  
  If we consider the $i$-th row of $A$, we obtain 
  $\mu^{(t+1)}_i=\sum_{j=1}^{n}m_j^{(t)}R_{i,j}(X)/Q_i(X)+c^{(t)}_i$. So under 
  the condition $c^{(t)}_i\in W_i$, $\mu^{(t+1)}_i$ can be expressed as a 
  rational fraction of the form $R'_i/Q_i$ and $\deg(R'_i)\leq 
  \max_j(\deg(R_{i,j}(X))$, this implies $c^{(t+1)} \in W_i$. 
\end{IEEEproof}

Following this result we want to limit the ``carries'' part of a RLM to the 
domain $W$. So we give the following definition for RLFSMs, which is a true
finite state machine.

\begin{definition}\label{def-rlfsm}
A Rational Linear Finite State Machine (RLFSM) with $\ell$-bit
output and length $n$ over $\Q$ is a finite state automaton defined by a 
pair $(A,C)$ of matrices over
 $\Q$ , with respective sizes $n\times n$ and $\ell\times n$. The space of 
 states of this automaton is $\F_2^n \times W$ where $W$ is defined from $A$ as 
 previously explained, the transition and extraction functions at time $t$ are
defined 
 by: if the automaton is in the state $(m^{(t)},c^{(t)})$ at time $t$ and 
 $v^{(t)}$ is the output at time $t$, then
\[
\left\{\begin{array}{rcl}
m^{(t+1)}&=&Am^{(t)}+c^{(t)} \bmod X\\
c^{(t+1)}&=&Am^{(t)}+c^{(t)} \Div X\\
v^{(t)}&=&Cm^{(t)}
\end{array}\right.
\]
\end{definition}

Now, we want to characterize in more details the output of a RLFSM. Set 
$G(X)=\prod_{i=1}^{n}Q_i(X)$. We have $A=\frac{1}{G(X)} A'$, where $A'$ is a
matrix 
with polynomial coefficients.

From the definition of $A'$, we have $\det(I-XA)=\frac{1}{G(X)^n}\det(G(X)I -
XA')$ 
where $T(X)=\det(G(X) I - X  A')$ is a polynomial. 
So we obtain $(I-XA)^{-1} =\frac{G(X)^n}{T(X)}\adj(I - XA')$, 
where $\adj(I - XA')$ is a matrix with 
polynomial coefficients.

We can easily deduce the rational form of the output of a RLFSM
\begin{proposition}
    Let ${\C L}$ be a RLFSM defined by a transition matrix $A$ and any output 
    matrix $C$. Set $T(X)=\det(G(X) (I - XA))$. The output sequences
$V^{(t)}_i$ are rational power 
    series of the form $P_i(X)/T(X)$.
\end{proposition}

\begin{IEEEproof}
  This result comes from the formula 
$$\begin{array}
{rcl}
   
  M^{(t)}  & = &   (I-X A)^{-1}(m^{(t)}+X  c^{(t)})\\   
    & = & \frac{G(X)^n}{T(X)}\adj(I - XA')(m^{(t)}+X c^{(t)}).   
\end{array}
$$ 
  Indeed, the denominators of the coefficients  of the matrix $(I-X 
  A)^{-1}$ are some divisors of $T(X)$, $m^{(t)}$ is a binary vector and 
  $c^{(t)}\in W$ is such that $G(X)^n X c^{(t)}$ is a polynomial vector.
\end{IEEEproof}

Note that the rational power series $P_i(X)/T(X)$ are {\it a priori} not
irreducible. 
In practice, the numerator is often the polynomial $Q(X)$ such that $Q(X)/P(X)$ 
is the irreducible rational representation of $\det(I-X A)$.
%%%%%%%%%%%%%%%%%%%%%%%%%%%%%%%%%%%%%%%%
\subsection{A first example}
We consider a filtered LFSR in Galois mode of size $n=12$ with connection 
polynomial $Q(X)=1+X^5+X^6+X^7+X^9+X^{11}+X^{12}$, filtered by a Boolean 
function in cells $m_0$, $m_5$, $m_7$ and $m_9$.

\begin{center}
\setlength{\unitlength}{3pt}
\begin{picture}(80,20)
% LFSR
\multiput(6,9)(6,0){12}{\framebox(3,6){}} 
\multiput(3,12)(6,0){12}{\vector(1,0){3}} 
\put(75,12){\line(1,0){3}}
\put(78,12){\line(0,1){6}}
\put(3,12){\line(0,1){6}}
\put(3,18){\line(1,0){75}}
\put(46,18){\vector(0,-1){6}}
\put(40,18){\vector(0,-1){6}}
\put(34,18){\vector(0,-1){6}}
\put(22,18){\vector(0,-1){6}}
\put(10,18){\vector(0,-1){6}}

% filtre

\put(19.5,9){\vector(0,-1){3}}
\put(31.5,9){\vector(0,-1){3}}
\put(43.5,9){\vector(0,-1){3}}
\put(73.5,9){\vector(0,-1){3}}
\put(18,6){\line(1,0){57}}
\put(18,6){\line(4,-1){8}}
\put(26,4){\line(1,0){41}}
\put(75,6){\line(-4,-1){8}}

\put(46.5,4){\line(0,-1){2}}
\put(46.5,2){\vector(-1,0){30}}
\put(6,1.2){output}

\end{picture}
\end{center}

If we are interested only on the filtered output bits, this LFSR can be
described by 
a RLFSM with the matrix 
$$
A'=\begin{pmatrix}
    X^4 & X^4 & 0 & 0  \\
    1+X & 0 & X & 0  \\
    X & 0 & 0 & X  \\
    X+X^2 & 0 & 0 & 0
\end{pmatrix}
$$

This matrix leads to a new representation of this RLFSM:

\begin{center}
\setlength{\unitlength}{3pt}
\begin{picture}(70,27)
% LFSR
\multiput(10,9)(16,0){4}{\framebox(6,8){}} 

\multiput(19,13)(16,0){3}{\circle{4}} 
\put(18,12.3){$\scriptscriptstyle{X}$}
\put(34,12.3){$\scriptscriptstyle{X}$}
\put(49.5,12.2){$\scriptscriptstyle{X^4}$}

\multiput(21,13)(16,0){3}{\vector(1,0){5}} 
\multiput(16,13)(16,0){3}{\line(1,0){1}} 
\put(64,13){\line(1,0){4}} 
\put(68,13){\line(0,1){12}} 
\put(7,25){\line(1,0){61}}

\multiput(7,20)(16,0){4}{\oval(8,4)} 
\put(3.7,19.3){$\scriptscriptstyle{X+X^2}$}
\put(22,19.3){$\scriptscriptstyle{X}$}
\put(36.3,19.3){$\scriptscriptstyle{1+X}$}
\put(54,19.3){$\scriptscriptstyle{X^4}$}
\multiput(7,25)(16,0){4}{\vector(0,-1){3}} 
\multiput(23,18)(16,0){3}{\vector(0,-1){5}} 
\put(7,18){\line(0,-1){5}}
\put(7,13){\vector(1,-0){3}}

% filtre

\multiput(13,9)(16,0){4}{\vector(0,-1){3}}
\put(10,6){\line(1,0){54}}
\put(10,6){\line(4,-1){8}}
\put(18,4){\line(1,0){38}}
\put(64,6){\line(-4,-1){8}}

\put(37,4){\line(0,-1){2}}
\put(37,2){\vector(-1,0){20}}
\put(6,1.2){output}

\put(11.3,12.5){$\scriptscriptstyle{m_{{}_{9}}}$}
\put(27.3,12.5){$\scriptscriptstyle{m_{{}_{7}}}$}
\put(43.3,12.5){$\scriptscriptstyle{m_{{}_{5}}}$}
\put(59.3,12.5){$\scriptscriptstyle{m_{{}_{0}}}$}
\end{picture}
\end{center}

Let $B=(I-X A')^{-1}$, and $Q(X)=\det(I-X A')= 
X^{12}+X^{11}+X^9+X^7+X^6+X^5+1$. Then the value of $B$ is

$$
B= \frac{1}{Q(X)}\begin{pmatrix}
    P_{1,1} & P_{1,2} & P_{1,3} & P_{1,4}  \\
    P_{2,1} & P_{2,2} & P_{2,3} & P_{2,4}  \\
    P_{3,1} & P_{3,2} & P_{3,3} & P_{3,4}  \\
    P_{4,1} & P_{4,2} & P_{4,3} & P_{4,4}  \\
\end{pmatrix}
$$
 with $P_{1,1}(X)=1$, $P_{1,2}(X)=X^5$, $P_{1,3}(X)=X^7$, $P_{1,2}(X)=X^9$,
  $P_{2,1}(X)=X^7+X^6+X^4+X^2+X$, $P_{2,2}(X)=X^5+1$,  
  $P_{2,3}(X)=X^7+X^6+X^5+1$, $P_{2,4}(X)= X^9+X^8+X^7+X^2$,
  $P_{3,1}(X)=X^5+X^4+X^2$, $P_{3,2}(X)=X^{10}+X^9+X^7$,  
  $P_{3,3}(X)=X^7+X^6+X^5+1$, $P_{3,4}(X)= X^9+X^8+X^7+X^2$,
$P_{4,1}(X)=X^3+X^2$, $P_{4,2}(X)=X^8+X^7$, $P_{4,3}(X)=X^{10}+X^9$ and \\ 
$P_{4,4}(X)=X^9+X^7+X^6+X^5+1$.

If we denote by $(a_0,\ldots,a_{12})$ the initial state at time $t=0$ of the 
binary LFSR, then, the initial state of our RLFSM is
$m^{(0)}=(a_0,a_5, a_7,a_9)$ and
$c^{(0)})=(a_1+a_2X+a_3X^2+a_4X^3,a_6,a_8,a_{10}+a_{11}X)$ and the 
sequences in output are \\[2mm]
$\frac{a_0P_{1,1}(X)+a_5P_{1,2}(X)+a_7P_{1,3}(X)+a_9P_{1,4}(X)}{Q(X)}$
 \\ \phantom{a} \hfill
+$\frac{(a_1+a_2X+a_3X^2+a_4X^3)X}{Q(X)},$
\\
$\frac{a_0P_{2,1}(X)+a_5P_{2,2}(X)+a_7P_{2,3}(X)+a_9P_{2,4}(X)+a_{6}X}{Q(X)}, $
\\
$\frac{a_0P_{3,1}(X)+a_5P_{3,2}(X)+a_7P_{3,3}(X)+a_9P_{3,4}(X)+a_{8}X}{Q(X)},$
\\
$\frac{a_0P_{4,1}(X)+a_5P_{4,2}(X)+a_7P_{4,3}(X)+a_9P_{4,4}(X)+(a_{10}+a_{11})X}{Q(X)}$.

%%%%%%%%%%%%%%%%%%%%%%%%%%%%%%%%%%%%%%%%
\subsection{Application to windmill LFSRs}

Windmill LFSRs can be defined as LFSMs with no input and several outputs.
They have been introduced in \cite{Smeets88} as a cyclic cascade connection of
$v\geq1$
LFSMs. Each of these LFSMs is called a vane of the windmill. The classical
representation of those LFSMs is the Fibonacci one. However, in the rest of
this section, we will show them using the  equivalent Galois representation
because it is more suitable for a better understanding.  Windmill LFSRs are
characterized by their feedback and feedforward connections. These feedback and
feedforward connections are identical for all vanes, but the lengths of the
LFSMs may be different as they can be shifted in different LFSMs.
Figure~\ref{fig-vane} presents a generic vane in Galois mode.

Windmill LFSRs were introduced to achieve parallel generation of sequences.
Consider a sequence $S=(s_n)_{n\in\G N}$. While a classical automaton outputs
$s_0$ at the first clock, $s_1$ at the second, and so on, a parallel automaton
outputs $v$ bits at each clock: $(s_0,s_1,\ldots,s_{v-1})$ at the first clock,
$(s_v,\ldots,s_{2v-1})$ at the second, etc. More precisely a parallel automaton
has $v$ outputs and products the sequences $S^i:=(s_{n v+i})_{n\in\G N}$
where $0\le i<v$. Note that our study focus on characterizing the sequences
$S^i$ and not the reconstructed sequence $S$.

\begin{figure}[!t]
\centering
\includegraphics[width=0.95\columnwidth]{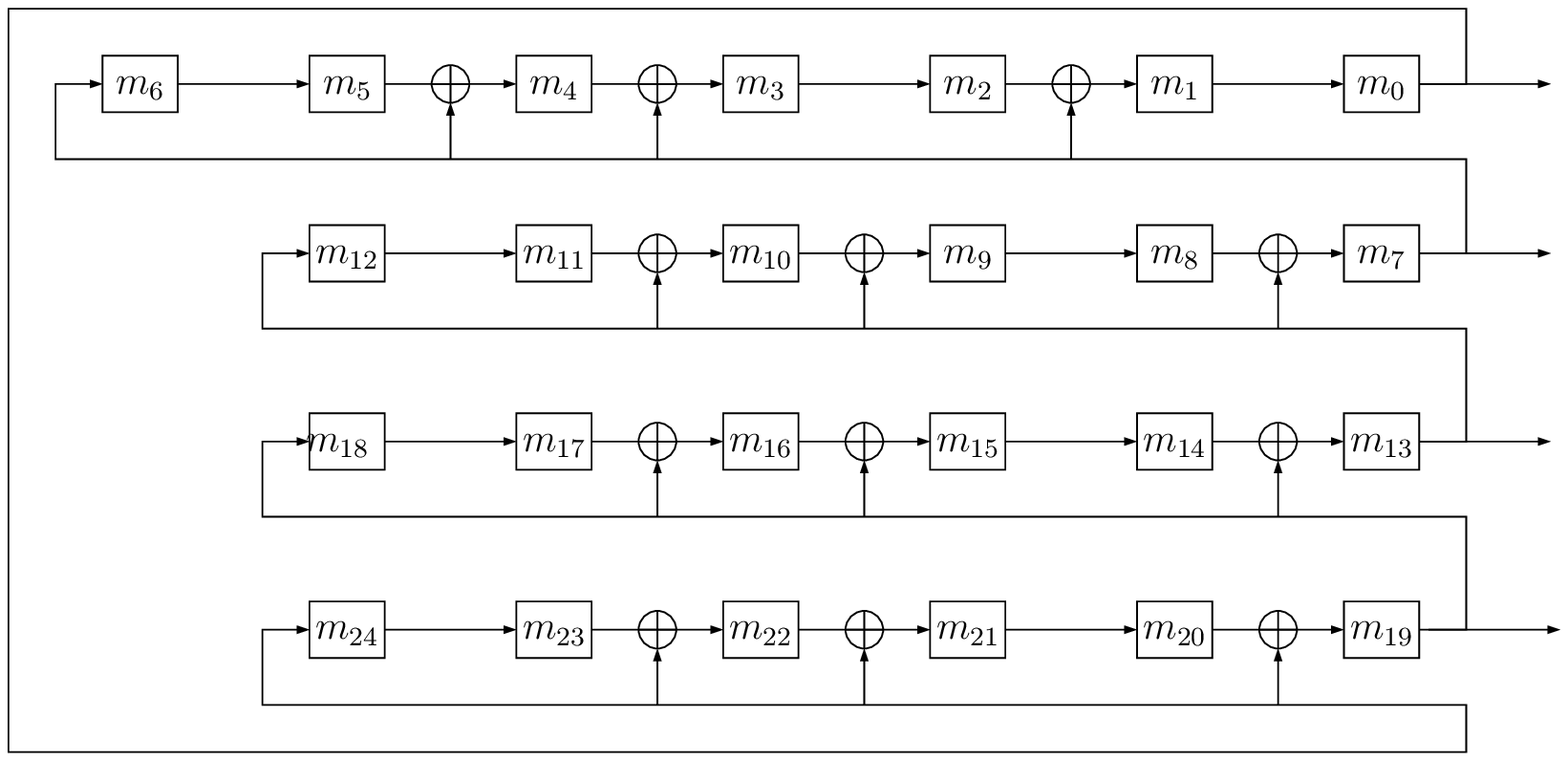}
\caption{A windmill with only feedforward connections.}\label{fig-windmill}
\end{figure}

Consider the windmill presented in Figure~\ref{fig-windmill} which is the one
used in
the stream cipher E0 \cite{E0}. It is constituted of one vane of length $7$ and
three identical vanes of length $6$. No feedback connection appears.
Feedforward connections appear, for example from cell $m_{13}$ to cells
$m_{12}$, $m_{10}$, $m_9$ and $m_7$.

\begin{figure}[!t]
\centering
\includegraphics[width=0.95\columnwidth]{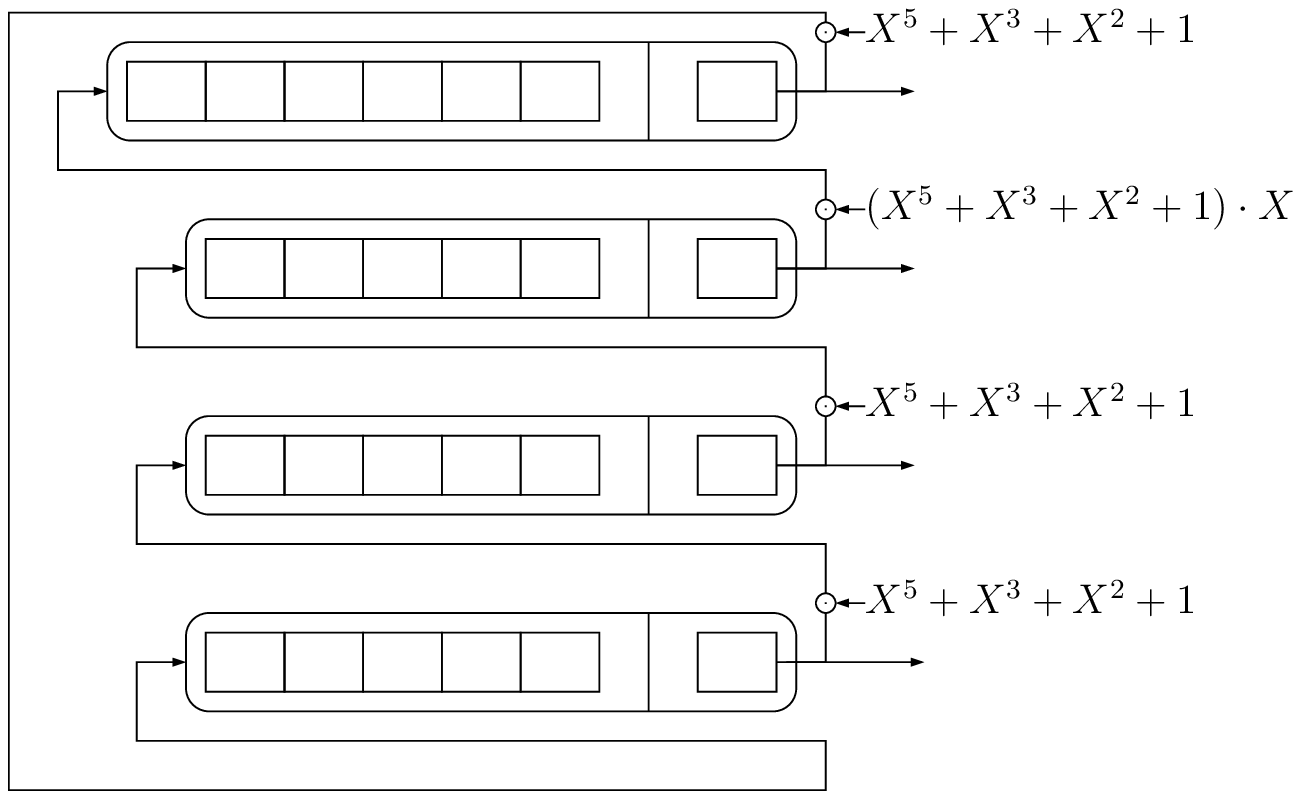}
\caption{A windmill in rational representation.}\label{fig-windmill2}
\end{figure}

Until now, only windmill LFSRs with a single vane repeated several times have
been studied. We generalize this definition allowing different vanes in a
windmill. We also give a new description of this windmill which will be more
compact. More precisely, using the example, we want to consider output
sequences of cells $m_0$,
$m_7$, $m_{13}$ and $m_{19}$, and characterize each vane by a polynomial. This
leads to the interpretation presented in Figure~\ref{fig-windmill2}.

With this definition the LFSM described in Figure~\ref{fig-windmill2} as the
following transition matrix:
\[
(X^5+X^3+X^2+1) \cdot
\begin{pmatrix}
0 & X & 0 & 0 \\
0 & 0 & 1 & 0 \\
0 & 0 & 0 & 1 \\
1 & 0 & 0 & 0
\end{pmatrix}
\]

We give in Table~\ref{tab-windmill2-clock} the values of $m^{(t)}$ and
$c^{(t)}$ during 8 clocks.
\begin{table*}
\[
\begin{array}{|c||c|c|c|c||c|c|c|c|}
\hline
Clock	&	m_0^{(t)}	&	m_1^{(t)}	&	
m_2^{(t)}	&	m_3^{(t)}	&	c_0^{(t)}	&	c_1^{(t)}
&	c_2'^{(t)}	&	c_3^{(t)}\\
\hline
\hline
0		&	1	&	0	&	0	&	0	&	0	&	0	&	0	&	0\\
1		&	0	&	0	&	0	&	1	&	0	&	0	&	0	&	X^4+X^2+X\\
2		&	0	&	0	&	1	&	0	&	0	&	0	&	X^4+X^2+X	&	X^3+X+1\\
3		&	0	&	1	&	0	&	1	&	0	&	X^4+X^2+X	&	X^3+X+1	&	X^2+1\\
4		&	0	&	0	&	0	&	1	&	X^5+X^3+X^2+1	&	X^3+X+1	&	X^4+X+1	&	X\\
5		&	1	&	1	&	0	&	0	&	X^4+X^2+X	&	X^2+1	&	X^4+X^3+X^2+X+1	&	1\\
6		&	0	&	1	&	1	&	0	&	X^5+X^2+X	&	X	&	X^3+X^2+X+1	&	X^4+X^2+X\\
7		&	0	&	1	&	1	&	0	&	X^5+X^4+X^3+X^2+X	&	X^4+X^2+X+1	&	X^2+X+1	&	X^3+X+1\\
8		&	0	&	0	&	1	&	1	&	X^5+X^4+X	&	X^4+X^3+X^2+1	&	X+1	&	X^2+1\\
\hline
\end{array}
\]
\caption{States of Figure~\ref{fig-windmill2} during 8
clocks.}\label{tab-windmill2-clock}
\end{table*}

According to Definition~\ref{def-rlfsm}, windmills as introduced by Smeets and
Chambers \cite{Smeets88} agree with the following definition:
\begin{definition}\label{def-windmill}
A windmill LFSR with polynomials $\alpha(X),\beta(X)$ with $\beta(0)\not=0$ and
$v$ vanes is an LFSR of length $v$ with matrix $A$ over $\G F_2[[X]]$ of the
form:
\[
\begin{pmatrix}
0&	\frac{\alpha(X)}{\beta(X)} X^{i_0}&	&(0)\\
\vdots&	\ddots&	\ddots&	\\
0&	(0)&	\ddots&	\frac{\alpha(X)}{\beta(X)} X^{i_{v-2}}\\
\frac{\alpha(X)}{\beta(X)} X^{i_{v-1}}&	0&	\dots&	0\\
\end{pmatrix}
\]
where $0\le i_0,\ldots,i_{v-1}$.
\end{definition}

With this representation each row represents a vane of the windmill. In
particular, as described in the following section the length of the vane $j$ is
equal to $\max(\deg(\alpha(X) X^{i_j}),\deg(\beta(X)))$.

By a straightforward calculus, we obtain $\det(I-X A)=
X^n\left(\alpha(X)/\beta(X)\right)^v+1$, where $n=i_0+\cdots+i_{v-1}$. Set 
$Q(X)=X^n\alpha(X)^v+\beta(X)^v$, it becomes $\det(I-X 
A)=Q(X)/\beta(X)^v$. The sequences $M_i^{(t)}$ observed in the output of this 
RLFSM are of the form $P_i(X)/Q(X)$. The main result on windmill generators 
(c.f. \cite{Smeets88}) is the fact that there exists a permutation $\sigma$ of 
$\{0,\ldots,v-1\}$ such that the series 
$S(X)=\sum_{t}(\sum_{i=0}^{v-1}m_i(t)X^{\sigma(i)})X^{vt}$ is a rational power 
series of the form $P(X)/Q(X^v)$. In other words, a windmill generator is able 
to output in parallel at each iteration $v$ consecutive values of a 
rational power series. The most interesting case is the one where $Q(X^v)$ is a 
primitive polynomial. Such windmill generators are used in the specification 
of the pseudo-random generator E0 included in the specifications of Bluetooth 
\cite{E0}.

Our polynomial approach gives a more synthetic point of view on these windmill 
generators. In particular, it shows that the windmill properties (i.e. the 
parallel generation of a given $m$-sequence) is independent of the 
implementation of the vanes. This implementation can be made with Fibonacci 
vanes as in the original version, or with Galois vanes as presented previously
or 
with  ring vanes with better diffusion delay as we will see in the next 
section.

%%%%%%%%%%%%%%%%%%%%%%%%%%%%%%%%%%%%%%%%
\subsection{Implementation of RLFSMs}

In our previous examples, the starting point was a binary circuit, or a RLFSM 
with a particular structure for its matrix. The converse problem is ``how 
to construct an efficient implementation from a given transition matrix $A$ of 
a RLFSM''.  We will show on two examples that this task is not so easy.

\subsubsection{A first example}

Consider the RLFSM $\C L^1$ defined by the following transition matrix:
\[
A=\begin{pmatrix}
\frac{X^2}{X^3+1}&\frac{X}{X^2+X+1}\\
1&0
\end{pmatrix}
\]
We compute $(I-X A)^{-1}$ to characterize the output sequences:
\[
(I-X A)^{-1}=\begin{pmatrix}
\frac{X^{3} + 1}{X^{4} + X^{3} + 1} & \frac{X^{3} + X^{2}}{X^{4} + X^{3} + 1} \\
\frac{X^{4} + X}{X^{4} + X^{3} + 1} & \frac{1}{X^{4} + X^{3} + 1}
\end{pmatrix}
\]

\begin{figure}[!t]
\centering
\includegraphics[width=0.95\columnwidth]{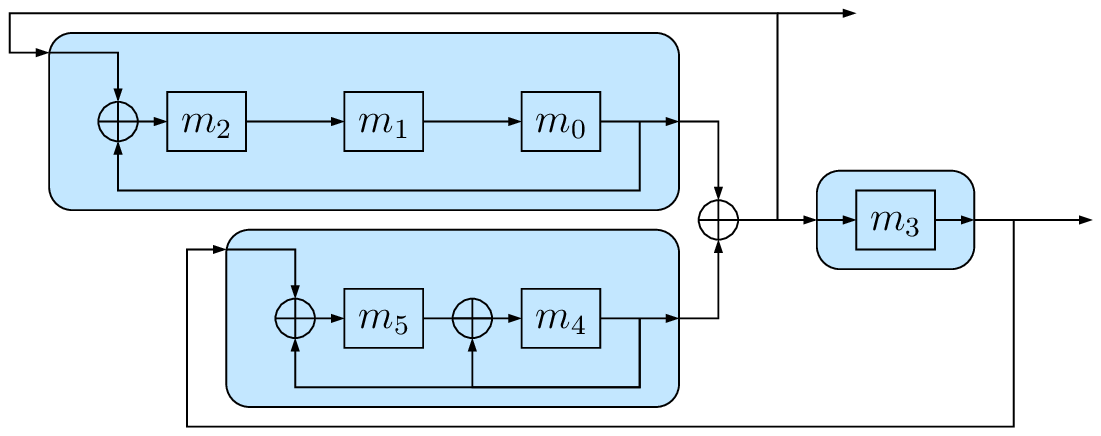}
\caption{First implementation of $\C L^1$.}\label{fig-ex1-1}
\end{figure}

Figure~\ref{fig-ex1-1} presents an implementation of this automaton built upon
three LFSMs. One for each nonzero coefficient in $A$.  These LFSMs are built
using a Galois vane architecture as presented in Figure~\ref{fig-vane}.

Note that, according to the notation of Figure~\ref{fig-ex1-1}, $\C L^1$ can be
expressed as the LFSM $(A',0,C')$ with:
\[
A'=\begin{pmatrix}
0&1&0&0&0&0\\
0&0&1&0&0&0\\
0&0&0&0&1&0\\
1&0&0&0&1&0\\
0&0&0&0&1&1\\
0&0&0&1&1&0
\end{pmatrix},\ C'=\begin{pmatrix}
1&0&0&0&1&0\\
0&0&0&1&0&0
\end{pmatrix}
\]

In particular, we have the following relations according to
Theorem~\ref{theo-lfsm}:
\begin{multline*}
V^{(t)}=\frac{1}{X^4+X^3+1}\quad \times\\
\begin{pmatrix}
1 & X & X^{2} & X^{3} + X^{2} & X + 1 & X^{2} + X \\
X & X^{2} & X^{3} & 1 & X^{2} + X & X^{3} + X^{2}
\end{pmatrix} m^{(t)}
\end{multline*}
This implementation is not optimal because it requires seven memories cells
while four are enough (it outputs sequences of the form $P(X)/(X^4+X^3+1)$ with
$\deg P(X)<3$). In particular, $\det(I-X A')=X^6+X^3+X^2+X+1$, i.e., this
automaton could output $m$-sequences of the form $P(X)/(X^6+X^3+X^2+X+1)$ using
a different matrix $C'$ because $X^6+X^3+X^2+X+1$ is primitive.

\begin{figure}[!t]
\centering
\includegraphics[width=0.95\columnwidth]{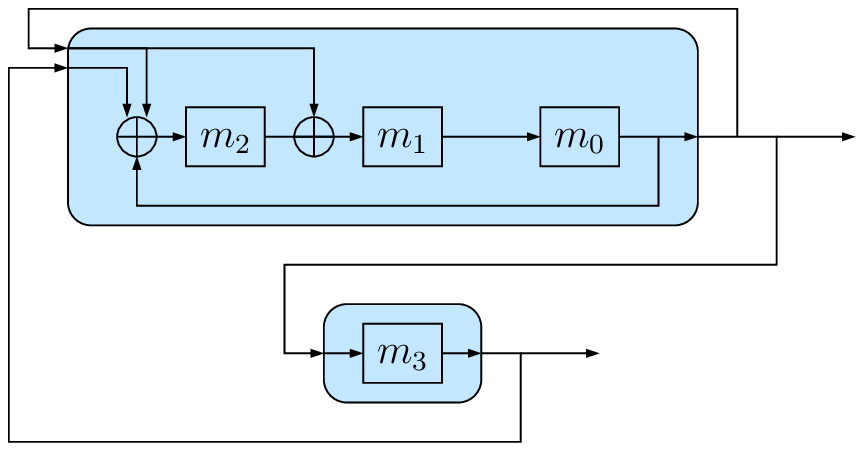}
\caption{Second implementation of $\C L^1$.}\label{fig-ex1-2}
\end{figure}

A better implementation is given considering one LFSM per line. To do so, note
that $\frac{X}{X^2+X+1}=\frac{X^2+X}{X^3+1}$. This leads to the implementation
presented in Figure~\ref{fig-ex1-2}.

As previously this leads to the relation:

\begin{multline*}
V^{(t)}= \\
\begin{pmatrix}
\frac{1}{X^{4} + X^{3} + 1} & \frac{X}{X^{4} + X^{3} + 1} & \frac{X^{2}}{X^{4}
+ X^{3} + 1} & \frac{X^{3} + X^{2}}{X^{4} + X^{3} + 1} \\
\frac{X}{X^{4} + X^{3} + 1} & \frac{X^{2}}{X^{4} + X^{3} + 1} &
\frac{X^{3}}{X^{4} + X^{3} + 1} & \frac{1}{X^{4} + X^{3} + 1}
\end{pmatrix} m^{(t)}.
\end{multline*}

\vspace{3mm}
%%%%%%%%%%%%%%%%%%%%%%%%%%%%%%%%%%%%%%%%
\subsubsection{Second example}

Consider the RLFSM $\C L^2$ defined by the following transition matrix:
\[
A=\begin{pmatrix}
\frac{X + 1}{X^{3} + X + 1} & \frac{X}{X^{2} + X + 1} & 0 \\
X^{3} + X^{2} & X^{2} & 1 \\
0 & \frac{X + 1}{X^{2} + X + 1} & 0
\end{pmatrix}
\]

\begin{figure}[!t]
\centering
\includegraphics[width=0.95\columnwidth]{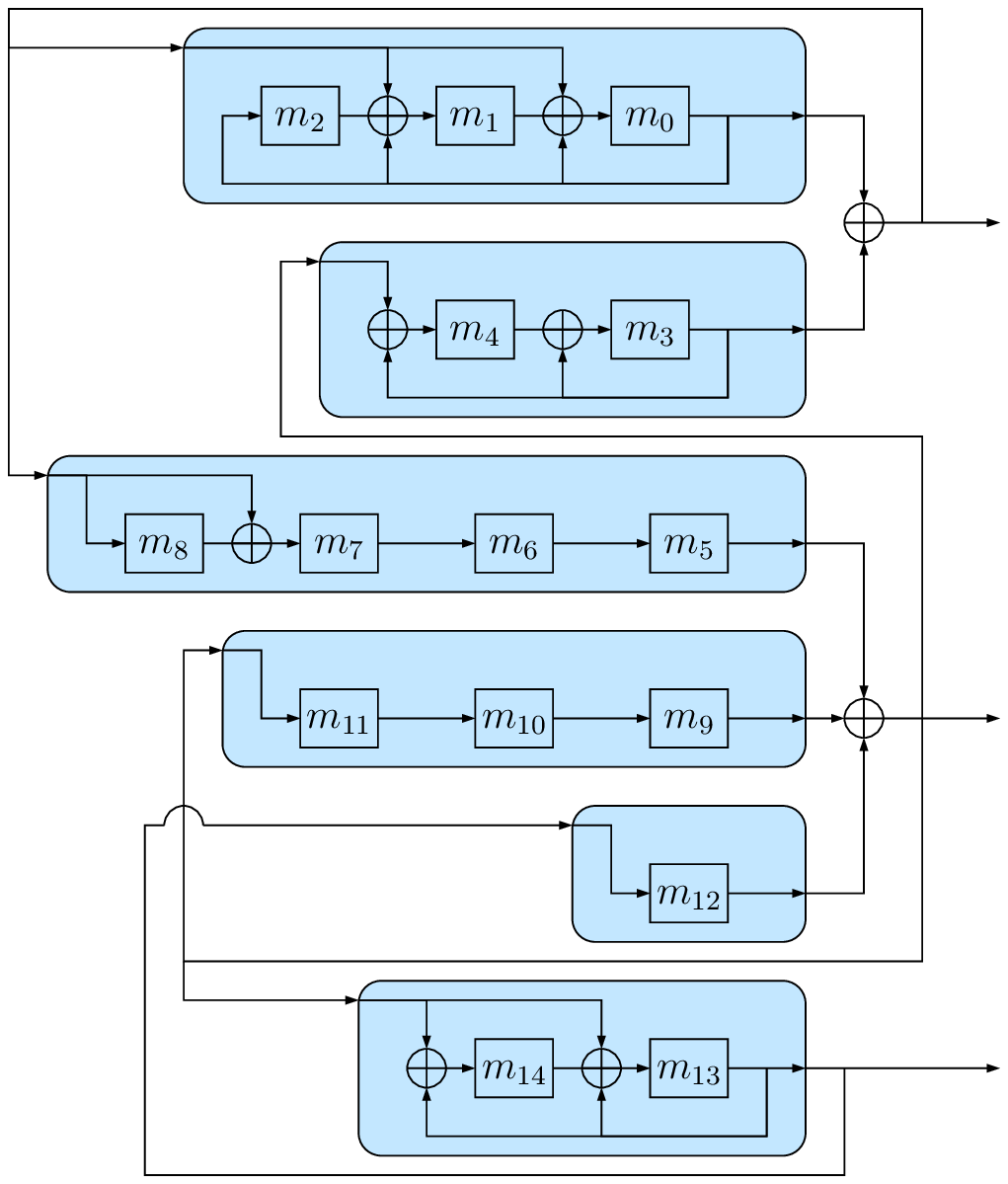}
\caption{First implementation of $\C L^2$.}\label{fig-ex2-1}
\end{figure}

Figure~\ref{fig-ex2-1} presents an implementation of this automaton built upon
six LFSMs. One for each nonzero coefficient in $A$. These LFSMs are built using
a Galois vane architecture as presented in Figure~\ref{fig-vane}.

Note that, according to the notation of Figure~\ref{fig-ex2-1}, $\C L^2$ can be
expressed as the LFSM $(A',0,C')$ with:
\begin{scriptsize}
\[
A'=\left(\begin{array}{ccccccccccccccc}
0 & 1 & 0 & 1 & 0 & 0 & 0 & 0 & 0 & 0 & 0 & 0 & 0 & 0 & 0 \\
1 & 0 & 1 & 1 & 0 & 0 & 0 & 0 & 0 & 0 & 0 & 0 & 0 & 0 & 0 \\
1 & 0 & 0 & 0 & 0 & 0 & 0 & 0 & 0 & 0 & 0 & 0 & 0 & 0 & 0 \\
0 & 0 & 0 & 1 & 1 & 1 & 0 & 0 & 0 & 1 & 0 & 0 & 1 & 0 & 0 \\
0 & 0 & 0 & 1 & 0 & 1 & 0 & 0 & 0 & 1 & 0 & 0 & 1 & 0 & 0 \\
0 & 0 & 0 & 0 & 0 & 0 & 1 & 0 & 0 & 0 & 0 & 0 & 0 & 0 & 0 \\
0 & 0 & 0 & 0 & 0 & 0 & 0 & 1 & 0 & 0 & 0 & 0 & 0 & 0 & 0 \\
1 & 0 & 0 & 1 & 0 & 0 & 0 & 0 & 1 & 0 & 0 & 0 & 0 & 0 & 0 \\
1 & 0 & 0 & 1 & 0 & 0 & 0 & 0 & 0 & 0 & 0 & 0 & 0 & 0 & 0 \\
0 & 0 & 0 & 0 & 0 & 0 & 0 & 0 & 0 & 0 & 1 & 0 & 0 & 0 & 0 \\
0 & 0 & 0 & 0 & 0 & 0 & 0 & 0 & 0 & 0 & 0 & 1 & 0 & 0 & 0 \\
0 & 0 & 0 & 0 & 0 & 1 & 0 & 0 & 0 & 1 & 0 & 0 & 1 & 0 & 0 \\
0 & 0 & 0 & 0 & 0 & 0 & 0 & 0 & 0 & 0 & 0 & 0 & 0 & 1 & 0 \\
0 & 0 & 0 & 0 & 0 & 1 & 0 & 0 & 0 & 1 & 0 & 0 & 1 & 1 & 1 \\
0 & 0 & 0 & 0 & 0 & 1 & 0 & 0 & 0 & 1 & 0 & 0 & 1 & 1 & 0
\end{array}\right)\]
\end{scriptsize}
and
\begin{scriptsize}
\[
C'=\left(\begin{array}{ccccccccccccccc}
1 & 0 & 0 & 1 & 0 & 0 & 0 & 0 & 0 & 0 & 0 & 0 & 0 & 0 & 0 \\
0 & 0 & 0 & 0 & 0 & 1 & 0 & 0 & 0 & 1 & 0 & 0 & 1 & 0 & 0 \\
0 & 0 & 0 & 0 & 0 & 0 & 0 & 0 & 0 & 0 & 0 & 0 & 0 & 1 & 0
\end{array}\right)
\]
\end{scriptsize}
This implementation is not optimal because it requires fifteen memories cells
while nine are enough because ${\deg(\det(I-X A))}=9$. In particular,
$\deg(\det(I-X A'))=11$.

\begin{figure}[!t]
\centering
\includegraphics[width=0.95\columnwidth]{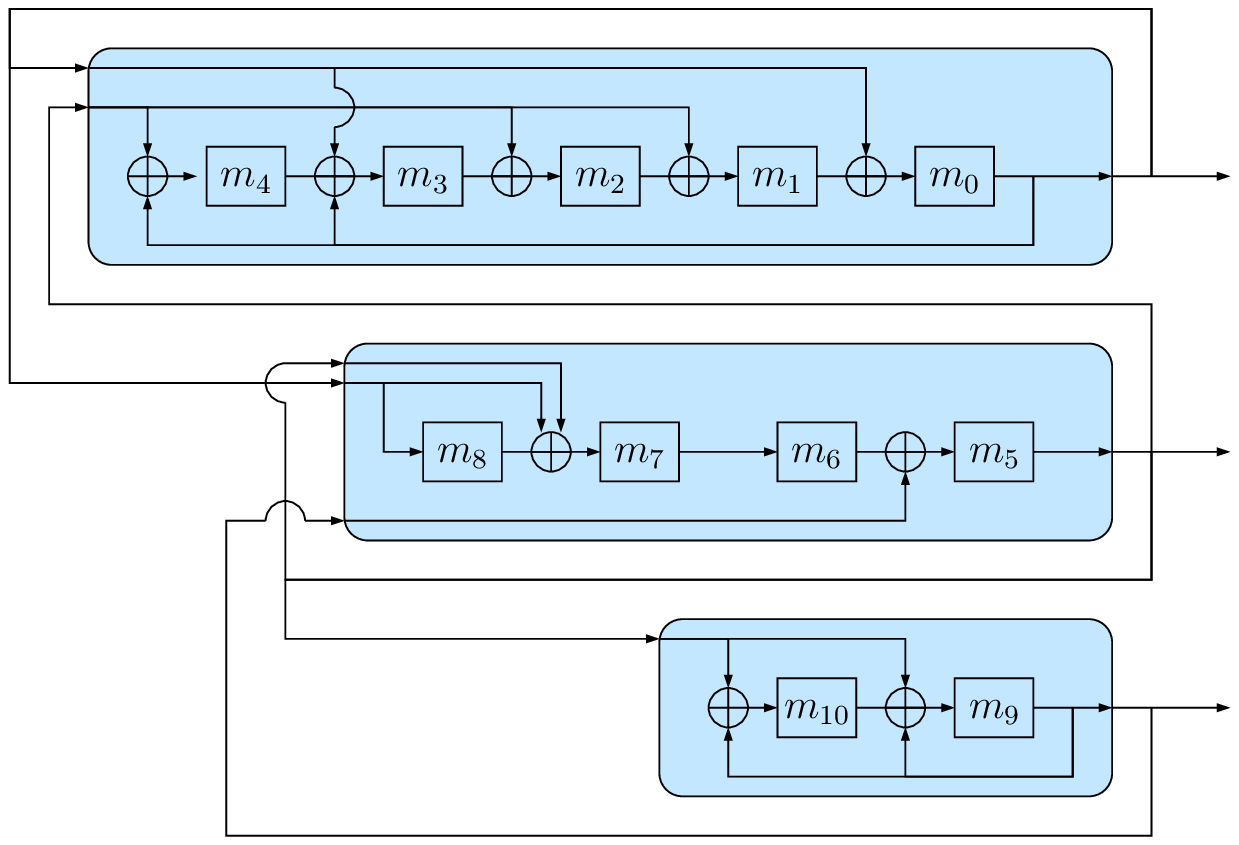}
\caption{Second implementation of $\C L^2$.}\label{fig-ex2-2}
\end{figure}

A better implementation is given considering one LFSM per line. This leads to
the implementation presented in Figure~\ref{fig-ex2-2}.

This implementation is still not optimum because it requires eleven memory
cells. This comes from the fact that in the matrix $A$, two terms with identical
denominator appears in the same column: $\frac{X}{X^2+X+1}$ and
$\frac{X+1}{X^2+X+1}$. More precisely, $\det(I-XA')=(X+1)(X^2+X+1)
(X^8+X^7+X^5+X^4+X^3+X^2+1)$. Thus, the automaton could be implemented
using the nine cells equivalent with the polynomial $(X+1)
(X^8+X^7+X^5+X^4+X^3+X^2+1)$ which is reducible and thus not primitive whereas
the last factor disappears inside the automaton itself.

\subsection{A practical example of application}

The rational representation is a theoretical tool that provides a global view
on the LFSRs design,
as seen for the case of windmill generators. However, previous examples have
shown that starting
from a circuit under rational representation to obtain an optimal 
implementation is not a simple task.

In the example given here, we generalize the windmills generators through
particular series circuits.
We limit our study with an example built on 3 circuits but the generalization
of this method is straightforward.

Let $A_1(X)=P_1(X)/Q_1(X)$, $A_2(X)=P_2(X)/Q_2(X)$ and $A_3(X)=P_3(X)/Q_3(X)$ 
be 3 elements of $\Q$. We consider the rational LFSR with transition matrix

$$T=\left(
\begin{array}
{ccc}
   
   0 & A_1 & 0  \\
   
   0 & 0 & A_2  \\
   
   A_3 & 0 & 0  \\

\end{array}
\right).$$
We have  $\det(I-XT)=1-X^3A_1A_2A_3=Q(X)/(Q_1(X)Q_2(X)Q_3(X))$ with 
$Q(X)=Q_1(X)Q_2(X)Q_3(X)+X^3Q_1(X)P_1(X)P_2(X)P_3(X)$.  
 The associated automaton 
computes rational series of the form $P(X)/Q(X)$.

Following the examples introduced in Figure \ref{fig-hard-ex} and in
Section \ref{1to1}, we choose $A_1(X)=A_2(X)=(X^6+X^3+1)/(X^8+X^6+X^5+X^3+1)$ 
and $A_3(X)=(X^7+X^5+X^4+X^2)/(X^8+X^6+X^5+X^3+1)$. The connection polynomial 
(i.e. the numerator of $\det(I-XT)$) is 
$Q(X)=X^{24} + X^{21} + X^{16} + X^9 + X^7 + X^3 + 1$. This 
polynomial is primitive, so the automaton will produce $m$-sequences.

For a practical implementation, we can replace the Galois vanes associated to 
$A_1(X)$, $A_2(X)$ and $A_3(X)$ by the ring vanes presented in Section 
\ref{1to1}.

This leads to a classical binary LFSR with transition matrix 

$$T_{r}=\left(
\begin{array}
{ccc}
   
   T_2& 0 & E_{1,4}  \\
   
   E_{1,1} & T_2 & 0 \\
   
   0 & E_{3,1}  & T_2  \\

\end{array}
\right).$$

Where $T_2$ is the $8\times 8$ matrix of the ring LFSR given in Figure 
\ref{fig-trans-ex} and where $E_{i,j}$ is the $8\times 8$ matrix with only one
1 in 
position $(i,j)$. The matrices $E_{i,j}$ represent the connections between the 3 
circuits. For example, the matrix $ E_{1,4}$ corresponds to the input 1 of the 
first ring LFSR and the output 4 of the third LFSR.

Note that $\det(I-XT_r)=Q(X)=X^{24} + X^{21} + X^{16} + X^9 + X^7 + X^3 + 1$.

Suppose now that we prefer an implementation with Galois vanes as internal 
blocks. The matrix of the Galois vane is the matrix $T_{0}$ given in Figure 
\ref{fig-trans-ex}. The multiplication by $A_1(X)$ is performed using 
${}^tB_1=(1,0,0,1,0,0,1,0)$ in input and $C_{1}=(1,0,0,0,0,0,0,0)$ for output. 
In the same way, we obtain $B_2=B_{1}$, ${}^tB_3=(0,0,1,0,1,1,0,1)$ and 
$C_{2}=C_{3}=C_{1}$. So the 
equivalent binary circuit is then

$$T_{g}=\left(
\begin{array}
{ccc}
   
   T_0& 0 & B_{1}C_{3}  \\
   
   B_{2}C_{1} & T_0 & 0 \\
   
   0 & B_{3}C_{2}  & T_0  \\

\end{array}
\right).$$

As we will see in the next section the automaton corresponding to the matrix 
$T_r$ has many nice properties 
compared to the classical ones obtained from the Galois LFSR. In particular, it 
needs 9 connections compared to 19 for the second one.

This example shows that the rational representation allows to separate the 
global design of the automaton from the choices of the hardware (or software) 
implementation.

The method presented in this example can be directly generalized to all 
Windmill generators and potentially leads to better practical implementations.

%%%%%%%%%%%%%%%%%%%%%%%%%%%%%%%%%%%%%%%%

\section{Design of efficient LFSRs for both hardware and software cryptographic
applications}\label{sec-implem}
In this section, we specialize our work on autonomous LFSMs, in particular on
LFSRs and their dedicated use for cryptographic applications. 

A general purpose of cryptography is to design primitives that are both
efficient in hardware and software because such primitives must run on all
possible supports, from RFID tags to super-calculators. Thus, cryptographers
must keep in mind, when they design cryptosystems, the very wide range of
targets on which cryptosystems must be rapid and efficient. As proof, the
Rijndael algorithm chosen as the AES \cite{DR02} in 2001 was one of the more
efficient algorithm in hardware and in software among the finalists of the AES
competition.

Thus, designing well-chosen dedicated LFSMs efficient both in hardware and in
software has direct consequences on the celerity of the cryptosystems which use
such primitives as building blocks. Among cryptographic primitives that use
LFSMs, we could cite the most famous case: the stream ciphers. Many stream
ciphers - such as E0 \cite{E0}, SNOW \cite{EJ02} or the finalists SOSEMANUK
\cite{DBLP:series/lncs/BerbainBCCGGGGLMPS08} and Grain v1
\cite{DBLP:series/lncs/HellJMM08} of the eStream project \cite{eStream} -
filter the content of one or many LFSMs to output pseudo-random bits. LFSMs
could
also be used as diffusion layer of a block cipher as proposed in \cite{COS}.
More recently, in \cite{Quark}, a particular LFSM combined with two NLFSRs
(Non-Linear Feedback Shift Registers) has been proposed at CHES 2010 as the
building block of a lightweight hash function named Quark. Well designing LFSMs
with good criteria is therefore crucial for symmetric key cryptography.

In this section, we first introduce the required design criteria that must be
fulfilled by an LFSM when used in cryptographic applications. We then extend the
traditional concept of diffusion (well-known in the block cipher context) to
the case of LFSMs. This leads to define a new criterion for good LFSMs choices
for cryptographic applications which is defined as the counterpart of the
Shannon diffusion concept \cite{shannon-otp}. 

Then, we present previous works on LFSMs for hardware and software cryptographic
applications. These automata have been widely studied
\cite{Golomb,Roggeman89,Smeets88,Klapper,Ecuyer05,Lau09a} and practical
constructions have emerged. We finally propose an efficient construction
dedicated to 
hardware and a second one dedicated to
software. This software construction is also efficient in hardware.

%%%%%%%%%%%%%%%%%%%%%%%%%%%%%%%%%%%%%%%%
\subsection{Design criteria}
We focus our design analysis on two important properties. The first one
characterizes the kind of sequences that are required for cryptographic
applications whereas the second one tries to formalize the notion of diffusion delay
in the context of LFSRs.

\subsubsection{$m$-sequences}

As introduced in Section~\ref{sec-background}, $m$-sequences are particular
linear recurring sequences with good properties \cite{Golomb,Klapper}. For
example, we give some properties for $m$-sequences of degree $n$ over $\G F_2$:
\begin{itemize}
\item an $m$-sequence is balanced: the number of $1$ is one greater than the
number of $0$ (considering one period).
\item an $m$-sequence has the run property: a run is a sub-sequence of $1$ or
$0$ followed and followed by 0 or 1. Half of the runs are of length $1$, a
quarter of length $2$, an eighth of
length $3$, etc. up to the 1-run of length $n$.
\item an $m$-sequence is a punctured De Bruijn sequence.
\item an $m$-sequence has the (ideal) two-level autocorrelation function where
the autocorrelation function for a binary sequence $a$ is defined as
$C_a(\tau) = \sum_{i=0}^{N-1} (-1)^{a_{i+\tau}+a_i}$ where $N$ is the period of
the sequence. This function verifies for a $m$-sequence:
$C_{\tau} = N$ if $\tau = 0 \mod N$ and $C_{\tau} = K$ if $\tau \neq 0 \mod N$
(where $K$ is a constant equal to $-1$ if $N$ is odd and to 0 is $N$ even).
\item an $m$-sequence has maximum period: an $m$-sequence verifying a linear
relation of degree $n$ has a period of $2^n-1$.
\end{itemize}

In the sequel, we are specially interested in LFSMs having a primitive 
connection polynomial and producing $m$-sequence which are the ones classically
used in cryptography. 
In particular, all our examples satisfy this condition.  However, most of the
results remains true without this hypothesis.

\subsubsection{Diffusion delay}
The concept of diffusion for a cipher was introduced by C. Shannon in
\cite{shannon-otp} as the dissipating effect of the redundancy of the
statistical structure of a message $M$. This concept is directly linked with
the Avalanche effect defined by H. Feistel in \cite{Feistel1973} which is a
desirable property of cryptographic algorithms, typically block ciphers and
cryptographic hash functions. The Avalanche effect means that if an input is
changed slightly,  the corresponding output must change significantly. In the
case of block ciphers, such a small change in either the key or the plaintext
should cause a drastic change in the ciphertext. 

Two precise notions could be directly derived: the strict avalanche criterion
(SAC) 
and the bit independence criterion (BIC). The strict avalanche criterion (SAC)
is a generalization of the avalanche effect. It is satisfied if, whenever a
single input bit is complemented, each of the output bits changes with a 50\%
probability \cite{WebsterT85}. The bit independence criterion (BIC) states that
output bits $j$ and $k$ should change independently when any single input bit
$i$ is inverted, for all $i$, $j$ and $k$.

When focusing on $m$-sequences, the measure of diffusion capacity is usually
studied through the notions of correlation, auto-correlation and
cross-correlation (see \cite{Golomb:2004:SDG:983299} for more details). The
correlation of two binary $m$-sequences $\alpha = (a_1, \cdots a_n)$ and $\beta
= (b_1, \cdots b_n)$ is measured as $C(\alpha, \beta) = \frac{1}{n} (A-D)$
where $A$ is the number of times for $i$ from 1 to $n$, that $a_i$ and $b_i$
agree and $D$ is the number of times that $a_i$ and $b_i$ disagree. The
auto-correlation of a given binary sequence has already been defined in the
previous subsection. It represents the similarity between a sequence and its
phase shift. The cross-correlation is defined as $C_{\alpha, \beta}(\tau) =
\sum_{i=0}^N (-1)^{a_{i+\tau} + b_i}$ when $q=2$ for two periodic binary
sequences $\alpha$ of period $s$ and $\beta$ of period $t$ with
$N=\mbox{lcm}(s,t)$ (for the case $q>2$ the reader could refer to
\cite{Golomb:2004:SDG:983299}).

Thus, in this part, we introduce a slightly different definition of diffusion
of an LFSM to more precisely capture the behavior of the beginning of a
sequence. This
parameter measures the time needed to mix the content of the cells of an
automaton. It could be expressed as the minimal number of clocks needed such
that any memory cell has been influenced by any other.
\begin{definition}
Let $\C L=(A,0,C)$ be an LFSM. Denote by $G$ the graph defined by the adjacency
matrix $A^t$, i.e., if $a_{i,j}\not=0$ then there exists a directed edge from
vertex $j$ and to vertex $i$. The diffusion delay is equal to the diameter of $G$.
\end{definition}

This parameter does not focus on the output sequence of
an LFSM but on the sequences produced INSIDE the
register itself (i.e. we look at the sequences $(m_0(t), \cdots, m_{n-1}(t), \cdots)$) 
and thus is relied on the implementation of the
automaton.

In a general point of view, if we take a random graph with $n$ vertices, the
average value of its diffusion delay is $\sqrt{n}$ as shown in \cite{FlajoletO89}.
For a complete graph, the diffusion delay parameter is optimal and is equal to 1,
however complete graphs do not produce good sequences as the corresponding
determinant 
$\det(I-AX)$ (where $A$ is the matrix representation of the complete graph) is
equal to $X+1$ if $n$ is odd and $1$ otherwise and thus could not produce 
sufficiently large $m$-sequences. 
Moreover, for a complete graph, from the circuit point of view, as the matrix 
of such graph as $n^2$ non-zero terms, this  
means that the representation circuit has $n^2-n$ xors. In the same way, the 
required number of xors for a circuit representing a random graph is about $n^2/2$. 
But, for cryptographic applications with efficient implementations, we look at 
circuits with good properties and with about $n/2$ xors which correspond with 
matrices with a binary weight equal to $3n/2$. Thus, we are far from circuits 
of complete or random graphs. 

So, we want to limit our study on lowering the diffusion delay when considering large
$m$-sequences.
More precisely, our aim in this section is double: we want to propose LFSRs
that produce large $m$-sequences with an efficient implementation and
with a low diffusion delay.

Let us explain now why it is important in cryptographic context to lower
diffusion delay.  
This criterion aims at evaluating the speed needed to completely spread a
difference into the automaton. More precisely, when considering an LFSM of size
$n$ with a diffusion delay $\delta$. Replacing the content of a cell $m_i^{(t)}$ by
$m_i^{(t)}+1$ may influence any cell $m_j$ with $0\le j<n$ after $\delta$
clocks. It could also be expressed in terms of correlation: after $\delta$
clocks, the behavior of any cell is correlated with any other. More precisely,
consider the two following sequences: the first sequence $\alpha = (a_1,
\cdots, a_N)$ is a binary sequence of the states of the content of the register
of an LFSR initialized with an $n$-bit word $a_1$ (i.e. each element $a_i$ of
$\alpha$ is the content at time $i$ of the LFSR and is $n$-bit long).
The second sequence of same length $N$, $\beta = (b_1, \cdots, b_N)$, is
constructed in the same way with an initialization $b_1$ that differ from $a_1$
on a single bit position. Then, $C(\alpha, \beta)$ is lowered by the LFSR with
the smaller diffusion delay for small values of $N$ (we have compared the results
obtained for three LFSRs of length $n=12$ bits (a Galois one, a Fibonacci one
and a Ring one) and correlation values until $N=256$). Note that the effect of
a small diffusion delay could only be observed for small values of $N$ because after
more clocks the influence of each modified bit is complete whatever the value
of the diffusion delay of the considered LFSR.

For example, considering Galois, Fibonacci LFSRs and Cellular automata of size
$n$, the associated diffusion delay is $n-1$ because the cells on each side $m_0$ and
$m_{n-1}$ require $n-1$ clocks to mix together. In the other hand, Ring LFSRs
allow to lower this parameter as its associated graph is closer to a random
graph, and as the expected value of the diameter of a random graph with $n$
vertices is $\sqrt n$. Ring LFSRs achieve a better diffusion delay. However, in
practice, this value is an average that could not be always reached especially
because we also focus our design choices on Ring LFSRs with sparse transition
matrix, i.e., we will consider graphs with few edges.

This diffusion delay criterion may be important for cryptographic purpose where small
differences in keys or in messages are required to have a large impact. It may
also be useful to lower the dimension gap for Pseudo Random Number Generators as
presented in \cite{Ecuyer96,Ecuyer05}. Hence, the dimension gap lowers when an
RNG outputs uniformly distributed point in a given sample space.

Moreover, this diffusion delay criterion could also be important, in stream cipher
design, to determine the number of clocks required by the so called
initialization phase and to speed up this step. Indeed, a stream cipher is
composed of two phases: an initialization phase where no bit are output and a
generation phase where bits are output. The initialization phase aims at mixing
together the key bits and the $IV$ bits. Thus, a lower diffusion delay allows to
speed up this mix in terms of number of clocks. 
For example, the F-FCSR v3 stream cipher proposed in
\cite{DBLP:conf/sacrypt/ArnaultBLMP09} based on a ring FCSR with a diffusion delay
equal to $d$ has an initialization phase with only $d+4$ clocks for mixing
purpose whereas the previous version of the F-FCSR family (F-FCSR v2) is based
on a Galois FCSR and thus requires $n+4$ clocks in the initialization step
where $n$ is the length of the considered FCSR. Thus, as $d<n$, a ring FCSR
with a ``good'' (i.e. low) diffusion delay allows to improve the general throughput
of the stream cipher by speeding up the initialization step.

As previously suggested by the example concerning FCSRs, because the diffusion delay 
criterion introduced in this section is essentially linked with the graph of the 
automaton whatever the considered graph, then the diffusion delay criterion could 
be applied for all possible automata: LFSRs, NLFSRs or FCSRs. For example, the FCSR 
used in the stream cipher F-FCSR v3 is a ring FCSR which has replaced a classical 
Galois FCSR. This modification leads to halve the number of required clocks 
during the initialization step 
and to completely discard the attack of Hell and Johannson \cite{HellJ08} 
against F-FCSR v2 due to a better internal diffusion delay.

%%%%%%%%%%%%%%%%%%%%%%%%%%%%%%%%%%%%%%%%
\subsection{Efficient hardware design}
We show in this subsection how to achieve good hardware design and we first
introduce the constraints required to achieve such a design:
\begin{itemize}
\item Critical path length: The shorter longest path must be as short as
possible to raise frequency.
\item Fan-out: A given signal should drive minimum gate number as exposed
in~\cite{Joux06}.
\item Cost: The number of logic gates must be as small as possible to lower
consumption.
\end{itemize}

We focus on these parameters because lowering these values allows to increase
the frequency of the automata, consequently it allows to increase the
throughput.

\subsubsection{Previous works}
Previous works have been done to lower those parameters. For example, in
\cite{Hybrid} the authors proposed top-bottom LFSR: a Ring LFSR divided in two
parts: a Fibonacci part and a Galois part corresponding with a transition
matrix of the form:
\[
A=\begin{pmatrix}
g_1	&	1	&		&		&		&		&		&		&		\\
g_2	&		&	1	&		&		&		&	(0)	&		&		\\
\vdots&		&		&	\ddots&		&		&		&		&		\\
g_{i-1}&	&		&		&	\ddots&		&		&		&		\\
	&		&		&		&		&	\ddots&		&		&		\\
	&		&	(0)	&		&		&		&	\ddots&		&		\\
	&		&		&		&		&		&		&	1	&		\\
1	&		&		&		&		&	f_i	&	f_{i+1}&\dots&	f_{n}
\end{pmatrix}
\]
This approach is a trade-off between Galois and Fibonacci LFSRs. In particular,
given a polynomial, there exists a top-bottom LFSR with this connection
polynomial. The critical path length, the fan-out and the cost may thus be an
average
between the Galois and the Fibonacci cases. But this construction also carries
the disadvantages of both cases, for example a slow diffusion delay.

%%%%%%%%%%%%%%%%%%%%%%%%%%%%%%%%%%
In \cite{Polonais}, the authors proposed a method that constructs, from a given
LFSR,
a similar LFSR with a lower critical path length and a lower fan-out. To do
so, they modify step by step the transition matrix of the original LFSR using
left and right shifts without modifying the corresponding value of the
connection
polynomial. For a given connection polynomial, those constructions lead to
implementations with a critical path of length at most $2$, a fan-out of at
most $3$
and a constant cost when starting the algorithm using a Galois LFSR. More
precisely, their method behaves well on polynomials with uniformly distributed
coefficients, i.e., polynomials with the same separation between any two
consecutive non-zero coefficients. They give as an example the polynomial
$X^{72}+X^{64}+X^{55}+X^{45}+X^{37}+X^{27}+X^{18}+X^9+1$, compared to
$X^{72}+X^{49}+X^6+X^5+X^4+X^3+X^2+X^1+1$. In summary, their method leads to
consider Ring LFSRs with transition matrix of the form
\[
A=\begin{pmatrix}
	&	1	&		&		&		&		&		&		&		\\
	&		&	1	&		&		&		&	(0)	&		&		\\
	&		&		&	\ddots	&		&		&		&		&		\\
	&		&		&		&	1	&		&		&		&		\\
(0)	&		&		&		&	h_1	&	1	&		&		&		\\
	&		&		&	\udots	&	h_2	&		&	\ddots	&		&		\\
	&		&	h_{n-4}	&	\udots	&		&		&		&	\ddots	&		\\
	&	h_{n-2}	&	h_{n-3}	&		&		&		&		&		&	1	\\
1	&	h_{n-1}	&		&		&	(0)	&		&		&		&	
\end{pmatrix}
\]
for the connection polynomial $X^n+h_{n-1}X^{n-1}+\dots+h_1X+1$ and $n$ odd (the
form is similar for $n$ even).

The authors also give a generic method (using two other elementary
transformations called SDL and SDR that preserve the connection polynomial) to
lower the hardware cost of an LFSR. To reach an LFSR with a better cost, the
authors must apply their method step by step until a x-or operation is reached
using their algorithm. The point of view taken in this article is thus from a
given connection polynomial and a given transition matrix to reach a better form
of the transition matrix (and thus a better hardware implementation) keeping
the same connection polynomial. The proposed methods are based on looking at
similar LFSRs. However, from a given LFSR, all the possible similar LFSRs could
not be reached using their algorithms. The corresponding diffusion delay of this kind
of LFSRs is about $n/2$. We show in the different examples given in this
Section that we could reach a better diffusion delay jointly with a more compact
implementation.

\subsubsection{Our approach}
Moreover, in most of the applications, the designer does not care about which
connection polynomial is chosen for the LFSR but only needs to know that the
connection polynomial is primitive. This is the core of our approach and of our
proposal where we randomly pick transition matrices with desired properties
(that could be application-dependent) and a posteriori verify if the obtained
connection polynomial is primitive or not. To do so, we first need to express
the
previous required constraints relying on the transition matrix of a Ring LFSR.
Table~\ref{tab-hard-design} sums up those constraints using the following
notations: denote by $\C L$ a Ring LFSR of length $n$ with transition matrix
$A$. We compute its connection polynomial $Q(X)$ and consider the associated
Galois LFSR $\C L_G$ and Fibonacci LFSR $\C L_F$. We denote by
$\col_0,\dots,\col_{n-1}$ the columns of~$A$ and $\row_0,\dots,\row_{n-1}$ its
rows. We note $w:=w_H(Q(X))$. All the presented constraints will be taken into
account in our approach in order to reach an LFSM that satisfies all the
requirements. 

\begin{table*}
\[
\begin{array}{|c||c|c|c|c|c|}
\hline
						&	\text{Galois}	&	\text{Fibonacci}		&	\text{Cellular automaton}	&
\text{Ring LFSR} &	\text{LFSR of \cite{Polonais}} \\
\hline\hline
\text{Critical path}	&	1				&	\lceil\log_2(w-1)\rceil	&	2							&
\max\lceil\log_2(w_H(\row_i))\rceil  & 2 \\
\hline
\text{Fan-out}			&	w-1				&	2						&	3							&	\max w_H(\col_i) & 3\\
\hline
\text{Cost}				&	w-2				&	w-2						&	n							&	w_H(T)-n & w-2 \\
\hline
\text{Diffusion delay}		&	n-1				&	n-1						&	n-1							&	\le n-1 & n/2\\
\hline
\end{array}
\]
\caption{Critical path, fan-out, cost and diffusion delay of Galois LFSRs, Fibonacci
LFSRs, Cellular automata, generic Ring LFSRs and construction proposed in
\cite{Polonais}.}\label{tab-hard-design}
\end{table*}

Galois LFSRs are optimal for the critical path, while Fibonacci LFSRs are
optimal
for the fan-out. A Ring LFSR can be built to reach these two values. More
precisely a Ring LFSR with a Hamming weight of at most 2 for its columns and
its rows will have an optimal critical path and an optimal fan-out with a good
diffusion delay as summed up in Table~\ref{tab-hard-design}.

However, we do not have an algorithm that construct an LFSR with a given
connection polynomial, we just can pick random transition matrix with good
properties. Hence, as we allow the connection to be freely chosen, the
constructed matrices do not present any special form allowing to compute
efficiently the connection polynomial. Moreover, when considering LFSMs in
practice, the constraint on the connection polynomial is simply to be primitive,
not to have a particular value.

%%%%%%%%%%%%%%%%%%%%%%%%%%%%%%%%%%%%%%%%
\begin{figure}[!t]
\begin{algorithmic}
\REQUIRE{$n$ the length of the Ring LFSR to seek. $f\le n$ the number of
feedbacks to place.}
\ENSURE{A transition matrix $A$ with a critical path of length $1$, a fan-out
of $2$
and a cost of $f$ logic gates and such that its connection polynomial is
primitive of degree $n$.}

\REPEAT
	\STATE $A\leftarrow (a_{i,j})_{0\le i,j<n}\text{ with
}{a_{i,j}=\left\{\begin{array}{l}
		1\text{ if }j\equiv i+1\bmod n\\
		0\text{ otherwise}\end{array}\right.}$
		\WHILE{$w_H(A)<n+f$}
			\STATE $(i,j)\leftarrow Random([0,n]\times[0,n])$
			\IF{$w_H(\row_i)=1$ AND $w_H(\col_j)=1$}
				\STATE $a_{i,j}\leftarrow 1$
			\ENDIF
		\ENDWHILE
		\STATE $Q(X)\leftarrow\det(I-XA)$
\UNTIL{$Q(X)$ is primitive}
\RETURN $A$
\end{algorithmic}
\caption{Algorithm to pick randomly a Ring LFSR with a good hardware
design.}\label{algo-hard1}
\end{figure}
%%%%%%%%%%%%%%%%%%%%%%%%%%%%%%%%%%%%%%%%

\begin{figure}[!t]
\centering
\includegraphics[width=0.95\columnwidth]{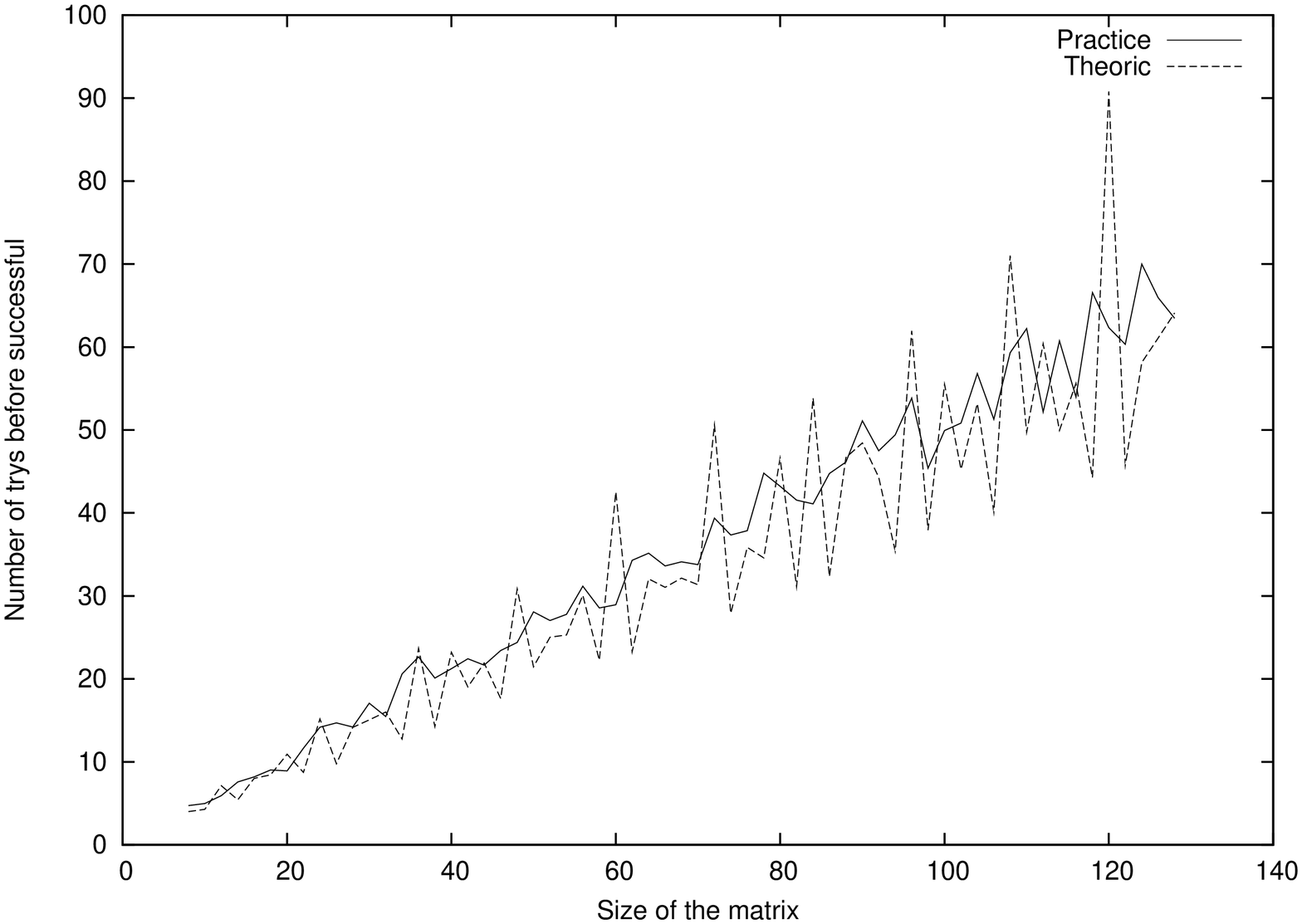}
\caption{Theoretic and empirical number of trials needed for
Algorithm~\ref{algo-hard1}.}\label{fig-complexity}
\end{figure}

Algorithm \ref{algo-hard1} picks random feedbacks positions and computes the
associated connection polynomial. This algorithm is probabilistic. We expect
picking a random matrix of size $n$ and computing its connection polynomial is
equivalent to pick a random polynomial of degree $n$. More precisely we know
that the connection polynomial as its constant coefficient and its greatest
coefficient equal to $1$, so the number of possibly constructed polynomials is
$2^{n-2}$. The number of primitive polynomials of degree $n$ over $\G F_2$ is
$\frac{\varphi(2^n-1)}{n}$ where $\varphi$ is the Euler function. We expect
Algorithm~\ref{algo-hard1} to be successful after
$\frac{2^{n-2}}{\varphi(2^n-1)/n}$ tries as presented in
Fig.~\ref{fig-complexity}.

The time complexity of this algorithm is driven by the time it takes to compute
$\det(I-XA)$ which is roughly $\C O(n^3)$.

%%%%%%%%%%%%%%%%%%%%%%%%%%%%%%%%%%%%%%%%
For a hardware oriented LFSM, each feedback can be freely placed. Using this
property we can lower the complexity of the previous algorithm using
intermediate computations done using the cofactors of the matrix $A$ as follows:
\begin{proposition}
Given a matrix $A$ over a ring $R$ of size $n\times n$. Note $E_{i,j}$ the
matrix with a single $1$ in position $i,j$. Then  we have $\det(A+\lambda
E_{i,j})=\det(A)+\lambda\cof_{i,j}$ where $\cof_{i,j}$ denotes the $(i,j)$-th
cofactor
of the matrix $A$.
\end{proposition}

The cofactors matrix of a matrix is equal to the transposition of its adjunct
matrix, which could be computed with classical inversion algorithms. Using the
previous proposition, we are able to improve the complexity of our algorithm
using Algorithm \ref{algo-hard2}.

%%%%%%%%%%%%%%%%%%%%%%%%%%%%%%%%%%%%%%%%
\begin{figure}[!t]
\begin{algorithmic}
\REQUIRE{$n$ the length of the Ring LFSR to seek. $f\le n$ the number of
feedbacks to be placed.}
\ENSURE{A transition matrix $A$ with a critical path of length $1$, a fan-out
of $2$
and a cost of $f$ logic gates and such that its connection polynomial is
primitive of degree $n$.}

\LOOP
		\STATE $A\leftarrow (a_{i,j})_{0\le i,j<n}\text{ with
}{a_{i,j}=\left\{\begin{array}{l}
		1\text{ if }j\equiv i+1\bmod n\\
		0\text{ otherwise}\end{array}\right.}$
		\WHILE{$w_H(A)<n+f-1$}
			\STATE $(i,j)\leftarrow Random([0,n]\times[0,n])$
			\IF{$w_H(\row_i)=1$ and $w_H(\col_j)=1$}
				\STATE $a_{i,j}\leftarrow 1$\;
			\ENDIF
		\ENDWHILE
		\STATE $C\leftarrow$ cofactors matrix of $I-XA$
		\STATE $Q_0(X)\leftarrow\det(I-XA)$
		\FOR{$0\le i,j<n$}
			\IF{$w_H(\row_i)=1$ and $w_H(\col_j)=1$}
				\STATE $Q(X)\leftarrow Q_0(X)-XC_{i,j}$
				\IF{$Q(X)$ is primitive}
					\STATE Break
				\ENDIF
			\ENDIF
		\ENDFOR
\ENDLOOP
\RETURN $A$
\end{algorithmic}
\caption{Algorithm to pick randomly a Ring LFSR with a good hardware
design.}\label{algo-hard2}
\end{figure}
%%%%%%%%%%%%%%%%%%%%%%%%%%%%%%%%%%%%%%%%

The complexity of this algorithm is driven by the computation of the cofactors
matrix and its determinant which can be achieved by a common
algorithm. Each computation of cofactors matrix costs $\C O(n^3)$ operations.
With 
a single cofactors matrix, we test roughly $n^2-nf$
polynomials. So the average complexity is about $\C O(n)$ operations.

\subsubsection{Example}
We give in Appendix \ref{lfsr-alg1} an example of a hardware oriented LFSR of
length 128 found using Algorithm \ref{algo-hard2}. This LFSR has a primitive
connection polynomial which has an Hamming weight of 65. The diffusion delay of this
LFSR is only 27 whereas the corresponding diffusion delay for a Galois or a Fibonacci
LFSR would be 127.

%%%%%%%%%%%%%%%%%%%%%%%%%%%%%%%%%%%%%%%%
\subsection{Efficient software and hardware design}

In the previous subsection, we focus our work on an efficient algorithm to find
efficient LFSRs for hardware design. In this subsection, we will show how we
could adapt those results for efficient software design of an LFSR and show how
this design is also efficient in hardware. The main difference between hardware
and software is the atomic data size. In hardware we operate on single bits,
whereas in software bits are natively packed in words such that working on
single bits is not natural and needs additional operations. The word size
depends on the architecture of the processor: 8 bits, 16 bits, 32 bits, 64 bits
or more. To benefit from this architecture we propose to use LFSRs acting on
words. Let us first summarize the previous works that have been done to
optimize software performances of LFSRs. Then, we introduce our construction
method to build LFSRs efficient in software and in hardware.

\subsubsection{Previous works}
Firstly, the Generalized Feedback Shift Registers were introduced in
\cite{Lewis73} to increase the throughput. The main idea here was to
parallelize $w$ Fibonacci LFSRs. More formally, the corresponding matrix of such
a construction is: 
\[
A=\begin{pmatrix}
0&		I_w&	&		&		&		\\
&		0&		I_w&	&		(0)&	\\
&		&		0&		I_w&	&		\\
\phantom{\vdots}&(0)&\phantom{\ddots}&\ddots&\ddots&\phantom{\ddots}\\
&		&		&		&		0&		I_w\\
I_w&	a_{n-2}I_w&\dots&	a_2I_w&	a_1I_w&	a_0I_w
\end{pmatrix}
\]

where $I_w$ represents the $w \times w$ identity matrix over $\G F_2$ and where
the $a_i$ for $i$ in $[0,..,n-2]$ are binary coefficients. The matrix $A$ could
be seen at bit level but also at $w$-bits word level, each bit of the $w$-bits
word is in fact one bit of the internal state of one Fibonacci LFSR among the
$w$ LFSRs. 

In \cite{Roggeman89}, Roggeman applied the previous definition to LFSRs to
obtain the Generalized Linear Feedback Shift Registers but in this case the
matrix $T$ is always defined at bit level. In 1992, Matsumoto in
\cite{Matsumoto92} generalized this last approach considering no more LFSR at
bit level but at vector bit level (called word). This representation is called
Twisted Generalized Feedback Shift Register whereas the same kind of
architecture was also described in \cite{Matsumoto98} and called the Mersenne
Twister. In those approaches, the considered LFSRs are in Fibonacci mode seen at
word level with a unique linear feedback. The corresponding matrices are of the
form:
\[
A=\begin{pmatrix}
0&		I_w&	&		&		&		\\
&		0&		I_w&	&		(0)&	\\
&		&		0&		I_w&	&		\\
\phantom{\vdots}&(0)&\phantom{\ddots}&\ddots&\ddots&\phantom{\ddots}\\
&		&		&		&		0&		I_w\\
I_w&	0&		0&		L	&	0&		0
\end{pmatrix}
\]
where $I_w$ represents the $w \times w$ identity matrix and where $L$ is a
$w\times w$ binary matrix. In this case, the matrix is defined over $\G F_2$
but could also be seen at $w$-bits word level. This is the first generalization
of LFSRs specially designed for software applications due to the word oriented
structure.

The last generalization was introduced in 1995 in \cite{Niederreiter95} with
the Multiple-Recursive Matrix Method and used in the Xorshift Generators
described in \cite{Marsaglia:2003:XR} and well studied in \cite{Ecuyer05}. In
this case, the used LFSRs are in Fibonacci mode with several linear feedbacks.
The matrix representation is: 
\[
A=\begin{pmatrix}
0&		I_w&	&		&		&		\\
&		0&		I_w&	&		(0)&	\\
&		&		0&		I_w&	&		\\
\phantom{\vdots}&(0)&\phantom{\ddots}&\ddots&\ddots&\phantom{\ddots}\\
&		&		&		&		0&		I_w\\
A_r&	A_{r-1}&A_{r-2}&\dots&	A_2&	A_1
\end{pmatrix}
\]
where $I_w$ is the identity matrix and where the matrices $A_i$ are software
efficient transformations such as right or left shifts at word level or word
rotation. The main advantage of this representation is its word-oriented
software efficiency but it also preserves all the good LFSRs properties if the
underlying polynomial is primitive. Moreover, using the special form of the
transition matrix, the connection polynomial is efficiently computed with the
formula $P(X)=\det\left(I+\sum_{j=1}^rX^jA_j\right)$.

A particular case of the Multiple-Recursive Matrix Method is studied in
\cite{Tsaban03}. The authors proposed to consider matrices $A_i$ of the form
$a_i\cdot T$ where $T$ is a square matrix of size $w$, and $a_i$ are scalar
elements. In this case, an algorithm to construct LFSMs with primitive
polynomials is given. This paper was the first to introduce efficient word-oriented LFSRs,
thus solving the challenge proposed by Bart Preneel in \cite{DBLP:conf/fse/Preneel94}.

An other way to construct software oriented LFSRs is to consider LFSRs over $\G
F_{2^w}$ as done in \cite{EJ02,DBLP:series/lncs/BerbainBCCGGGGLMPS08}. The SNOW
LFSR is given in Appendix \ref{lfsr-snow}. This interpretation allows to use
table-lookup optimization and gives good results. Those automata could be
interpreted as linear automata over $\G F_2$ because of the mapping $\G
F_{2^w}\to (\G F_2)^w$. In particular, they can be consider as a special case
of our proposal.

\subsubsection{Our proposal for building LFSRs efficient in software and in
hardware}

As for the hardware case our approach focuses on the construction of a software
oriented transition matrix. To do so, we will use transition matrices defined
by block. In the next algorithm, $A$ will define a block matrix, i.e., $A$ is
taken in $\C M_{n/k}(\C M_k(\G F_2))$ for a matrix of size $n$ divided in
blocks of size $k$ over $\G F_2$. When an LFSR is being defined by block, we
call it a word-LFSR.

Moreover we will use the right and left shift operations (denoted $\gg$ and
$\ll$) which are fast and implemented at word level. Given a word size $k$ we
define the matrix $L$ of left shift as the matrix $k\times k$ with ones on its
overdiagonal and zeros elsewhere. Similarly, the matrix $R$ of right shift is
defined as the matrix $k\times k$ with ones on its sub-diagonal and zeros
elsewhere, such that we have:
\begin{eqnarray*}
L\cdot(x_0,x_1,\dots,x_{k-1})^t&=&(x_1,\dots,x_{k-1},0)^t\\
R\cdot(x_0,x_1,\dots,x_{k-1})^t&=&(0,x_0,x_1,\dots,x_{k-2})^t
\end{eqnarray*}

Remark that LFSRs over $\G F_{2^w}$ can be expressed as word-LFSRs where used
operations are multiplications on $\G F_{2^w}$ seen as a space vector over $\G
F_2$, i.e., there exists a bijection between $\G F_{2^w}$ and $(\G F_2)^w$.

According to the previous discussion we propose Algorithm \ref{algo-soft} to
build efficient software LFSRs.

%%%%%%%%%%%%%%%%%%%%%%%%%%%%%%%%%%%%%%%%
\begin{figure}[!t]
\begin{algorithmic}
\REQUIRE{$k$ the word size. $n$ the length of the LFSR to seek with $k|n$.
$f\le n/k$ the number of word-feedbacks to place.}
\ENSURE{A transition matrix $A$ define by block with a cost of $f$ shift and
xor operations and such that its connection polynomial is primitive of degree
$n$.}

\REPEAT
	\STATE $A\leftarrow(a_{i,j})_{0\le i,j<n/k}$
	\STATE \hspace{1cm}$\text{ with }a_{i,j}=\left\{\begin{array}{l}
	I_k\text{ if }j\equiv i+1\bmod n/k\\
	0\text{ otherwise}\end{array}\right.$
	\STATE $From\leftarrow Random([0,n/k]^f)$
	\STATE $To\leftarrow Random([0,n/k]^f)$
	\STATE $Shift\leftarrow
Random\left(\big([-k/2,k/2]\setminus\{0\}\big)^f\right)$
	\FOR{$l\leftarrow 0$ to $f-1$}
		\STATE $a_{To[l],From[l]}\leftarrow a_{To[l],From[l]}$
		\STATE \hspace{2cm}$+\left\{\begin{array}{l}
			L^{Shift[l]}\text{ if }Shift[l]>0\\
			R^{-Shift[l]}\text{ otherwise}\end{array}\right.$
	\ENDFOR
	\STATE $Q(X)\leftarrow\det(I-XA)$
\UNTIL{$Q(X)$ is primitive}
\RETURN $A$
\end{algorithmic}
\caption{Algorithm to pick randomly an LFSR with a good software
design.}\label{algo-soft}
\end{figure}
%%%%%%%%%%%%%%%%%%%%%%%%%%%%%%%%%%%%%%%%

This algorithm picks random word-feedbacks positions and shift values, and
computes the associated connection polynomial. The complexity of this algorithm
is about the same than Algorithm~\ref{algo-hard1} because we have not been able
to use the block structure of the matrix to lower the determinant computation
complexity.

\subsubsection{Example}
\begin{figure}[!t]
\centering
\subfigure[Transition matrix]{
$A=\begin{pmatrix}
	&	I_8	&	R^1	&		&		\\
	&		&	I_8	&		&		\\
	&		&		&	I_8	&		\\
	&		&		&	L^3	&	I_8	\\
I_8	&	L^1	&		&		&		
\end{pmatrix}$
}
\subfigure[Representation]{
\includegraphics[width=0.95\columnwidth]{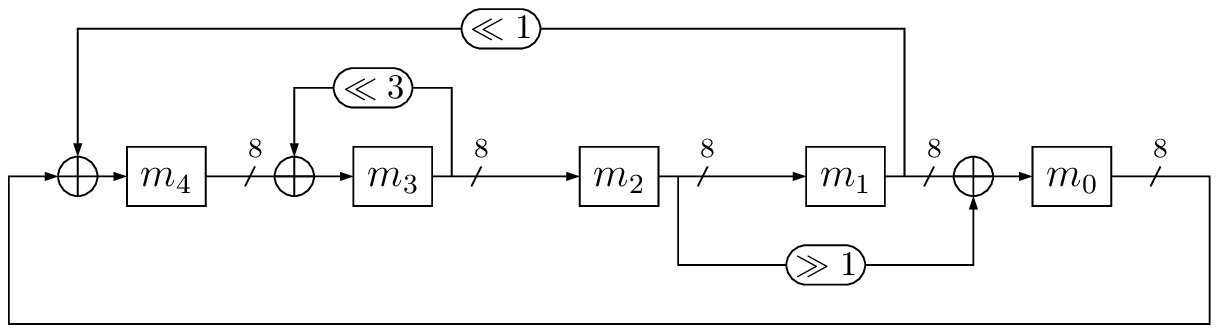}
}
\caption{An LFSR with efficient software design.}\label{fig-LFSR-para}
\end{figure}

We give in Figure~\ref{fig-LFSR-para} an example of an LFSR with an efficient
software design with $n=40$ and $k=8$ and a primitive connection polynomial. The
corresponding hardware implementation of this LFSR is also very good due to its
intrinsic structure (a fan out of 2, a critical path of length 1 and a cost of
19
adders) and because it fulfills the requirements of Alg \ref{algo-hard2}. The
diffusion delay of this LFSR is 27.

%%%%%%%%%%%%%%%%%%%%%%%%%%%%%%%%%%%%%%%%
Let us now also compare a word oriented LFSR picked using our algorithm to the
SNOW2.0 LFSR defined in \cite{EJ02}. The two LFSRs are respectively described in
Appendix \ref{lfsr-snow} and in Appendix \ref{lfsr-word}.

These two LFSRs output $m$-sequences of degree 512. We compare the diffusion delay and
the throughput in software for those two LFSRs:
\begin{itemize}
\item The diffusion delay of the SNOW LFSR is 49 compared to 33 for our LFSR.
\item The cost of one clock is 8 cycles for the SNOW LFSR using the sliding
window implementation as proposed in \cite{EJ02} (this technique could be only
applied for a Fibonacci LFSR). The cost for this LFSR implemented using
classical implementation is 20 cycles. The cost for our LFSR is 33 cycles.
\end{itemize}

As presented the diffusion delay is better for our LFSR. However, the cost of one
clock is higher in our case. This is due to the fact that the SNOW LFSR is
sparse (three feedbacks) while ours has 8 feedbacks. Moreover, the computations
are made using precomputed tables which leads to a better cost. However, the
hardware implementation of our own LFSR has a really low cost (it fulfills the
hardware design criteria we require in the previous section: critical path of 
length 1, fan-out of 2) whereas the SNOW2.0 LFSR could not be efficiently
implemented
in hardware due to the precomputed tables.

\subsection{Conclusion}
To sum up the results given in this section, we have proposed two algorithms
one for hardware purpose, one for software purpose that allow to build
efficient LFSRs with a low diffusion delay and good implementation criteria. Moreover,
building an LFSR using Alg. \ref{algo-soft} leads to an LFSR with good
cryptographic properties with an efficient implementation both in software and
in
hardware.

%%%%%%%%%%%%%%%%%%%%%%%%%%%%%%%%%%%%%%%%
\section{Conclusion}\label{sec-conclu}

In this paper, we have shown how to link together matrix representations and
polynomial representations for efficient LFSMs, LFSRs and windmill LFSRs
constructions. Those new representations lead to efficient implementations both
in software and in hardware. We have compared new Ring LFSR constructions with
LFSRs used in several stream ciphers and we have shown that Ring LFSRs have
always a better diffusion delay with better hardware performances and good software
performances. 

In further works, we aim at more precisely looking at the case of an LFSM with
$\ell$ output bits to give equivalent and general representations. We also want
to generalize those new results to Finite State Machines that are no more
linear. The same kind of generalization could be efficiently applied to
Feedback with Carry Shift Registers (FCSRs) or to Algebraic Feedback Shift
Registers (AFSRs).

%%%%%%%%%%%%%%%%%%%%%%%%%%%%%%%%%%%%%%%%%%%%%%%%%%%%%%%%%%%%%
\bibliographystyle{IEEEtran}
\bibliography{bib}
%%%%%%%%%%%%%%%%%%%%%%%%%%%%%%%%%%%%%%%%%%%%%%%%%%%%%%%%%%%%%

\newpage
\appendix

%%%%%%%%%%%%%%%%%%%%%%%%%%%%%%%%%%%%%%%%

\subsection{Example of a Ring LFSR of size 128 bits} \label{lfsr-alg1}

We describe a Ring LFSR of size 128 bits. The transition matrix $A=(a_{i,j})$
is given by:
\[
\left\{\begin{array}{l}
a_{i,i+1}=1\text{ for all }0\le i<127\\
a_{127,0}=1\\
a_{i,j}=1\text{ for }(i,j)\in\C F
\end{array}\right.
\]
where $\C F$ is the set:
\[
\left\{\begin{array}{cccccccc}
(4, 78),&(5, 19),&(8, 44),&(9, 106),\\
(10, 70),&(12, 14),&(14, 115),&(15, 55),\\
(17, 82),&(21, 64),&(22, 12),&(25, 127),\\
(27, 107),&(28, 112),&(31, 59),&(34, 111),\\
(35, 48),&(37, 36),&(38, 23),&(39, 88),\\
(43, 37),&(44, 26),&(46, 60),&(47, 100),\\
(49, 24),&(50, 25),&(51, 2),&(51, 27),\\
(55, 124),&(57, 113),&(59, 71),&(61, 29),\\
(69, 123),&(72, 52),&(73, 118),&(77, 46),\\
(80, 74),&(81, 83),&(83, 98),&(87, 53),\\
(88, 73),&(91, 47),&(93, 10),&(94, 21),\\
(95, 93),&(97, 13),&(98, 117),&(99, 50),\\
(100, 3),&(101, 104),&(104, 1),&(105, 114),\\
(106, 108),&(107, 105),&(109, 4),&(111, 28),\\
(112, 68),&(113, 42),&(114, 31),&(119, 18),\\
(120, 49),&(121, 32),&(123, 94),&(124, 6)
\end{array}\right\}
\]

This LFSR has a primitive connection polynomial. It has a cost of 64 adders, a
fan-out equal to 2 and a critical path of 1, and a diffusion delay of 27.

%%%%%%%%%%%%%%%%%%%%%%%%%%%%%%%%%%%%%%%%
\subsection{Description of the LFSR in SNOW 2.0 over $\G F_2$} \label{lfsr-snow}

We give here a description of the LFSR used in SNOW 2.0 \cite{EJ02} seen as a
LFSR over $\G F_2$.

First this LFSR is defined as a Fibonacci LFSR over $\G F_{2^{32}}$. The field
$\G F_{2^{32}}$ is defined as an extension of $\G F_{2^8}$ to allow an
efficient implementation and to prevent the guess-and-determine attack
presented in
\cite{Hawkes02}.

The implementation is based upon the multiplication by $\alpha\in\G F_{2^{32}}$
satisfying $\alpha\cdot
(c_3\alpha^3+c_2\alpha^2+c_1\alpha^1+c_0)=(c_2\alpha^3+c_1\alpha^2+c_0\alpha)+c_3\cdot
V$ with $V$ an element in $\G F_{2^{32}}$. We denote $M_\alpha$ the matrix of
this linear application seen over $\G F_2^{32}$:
\[
M_\alpha=\left(\begin{array}{ccc|c|c|c|c}
0	&0	&0	&	&&&\\
I_8	&	&	(0)&V_0&V_1&\dots&V_7\\
	&I_8&	&	&&&\\
(0)	&	&I_8&	&&&\\
\end{array}\right)\]
where 
\[
\left\{\begin{array}{c}
V_0={}^t(\texttt{0xE19FCF13})\\
V_1={}^t(\texttt{0x6B973726})\\
V_2={}^t(\texttt{0xD6876E4C})\\
V_3={}^t(\texttt{0x05A7DC98})\\
V_4={}^t(\texttt{0x0AE71199})\\
V_5={}^t(\texttt{0x1467229B})\\
V_6={}^t(\texttt{0x28CE449F})\\
V_7={}^t(\texttt{0x50358897})
\end{array}\right.
\]

Then the transition matrix of the LFSR of SNOW2.0 is presented in
Figure~\ref{snow-mat}.
\onecolumn

\begin{figure*}[!t]
\centering
\[
\left(\begin{array}{cccccccccccccccc}
	&	I_{32}&	&	&	&	&	&	&	&	&	&	&	&	&	&	\\
	&	&	I_{32}&	&	&	&	&	&	&	&	&	&	&	&	&	\\
	&	&	&	I_{32}&	&	&	&	&	&	&	&	&	&	&	&	\\
	&	&	&	&	I_{32}&	&	&	&	&	&	&(0)&	&	&	&	\\
	&	&	&	&	&	I_{32}&	&	&	&	&	&	&	&	&	&	\\
	&	&	&	&	&	&	I_{32}&	&	&	&	&	&	&	&	&	\\
	&	&	&	&	&	&	&	I_{32}&	&	&	&	&	&	&	&	\\
	&	&	&	&	&	&	&	&	I_{32}&	&	&	&	&	&	&	\\
	&	&	&	&	&	&	&	&	&	I_{32}&	&	&	&	&	&	\\
	&	&	&	&(0)&	&	&	&	&	&	I_{32}&	&	&	&	&	\\
	&	&	&	&	&	&	&	&	&	&	&	I_{32}&	&	&	&	\\
	&	&	&	&	&	&	&	&	&	&	&	&	I_{32}&	&	&	\\
	&	&	&	&	&	&	&	&	&	&	&	&	&	I_{32}&	&	\\
	&	&	&	&	&	&	&	&	&	&	&	&	&	&	I_{32}&	\\
	&	&	&	&	&	&	&	&	&	&	&	&	&	&	&	I_{32}\\
M_\alpha&	0&	I_{32}&	0&	0&	0&	0&	0&	0&	0&	0&	(M_\alpha)^{-1}&	0&	0&	0&	0\\
\end{array}\right)
\]
\caption{Transition matrix of SNOW2.0}\label{snow-mat}
\end{figure*}

%%%%%%%%%%%%%%%%%%%%%%%%%%%%%%%%%%%%%%%%
\subsection{Example of a word-oriented LFSR of size 512 bits} \label{lfsr-word}

We give in Figure~\ref{lfsr-word-mat} a description of a word-oriented LFSR of
length 512 with words of 32 bits. The grid in the matrix is drawn for
readability.

\begin{figure*}[!t]
\centering
\[
\left(\begin{array}{ccc|ccc|ccc|ccc|ccc|c}
	&	I_{32}&	&	&	&	&	&	&	&	&	&	&	&	&	&	\\
	&	&	I_{32}&	&	&	&	&	&	R^{14}&	&	&	&	&	&	&	\\
	&	&	&	I_{32}&	&	&	L^8&	&	&	&	&	&	&	&	&	\\
\hline
	&	&	&	&	I_{32}&	&	&	&	&	&	(0)&	L^{12}&	&	&	&	\\
	&	&	&	&	&	I_{32}&	&	&	&	&	&	&	&	&	&	\\
	&	&	&	&	&	L^2&	I_{32}&	&	&	&	&	&	&	&	&	\\
\hline
	&	&	&	&	&	&	&	I_{32}&	&	&	&	&	&	&	&	\\
	&	&	&	&	&	&	&	&	I_{32}&	&	&	&	&	&	R^{11}&	\\
L^{13}&	&	&	&	&	&	&	&	&	I_{32}&	&	&	&	&	&	\\
\hline
	&	&	&	&(0)&	&	&	&	&	&	I_{32}&	&	&	&	&	\\
	&	&	&	&	&	&	&	&	&	&	&	I_{32}&	&	&	&	\\
	&	&	&	&	&	&	&	&	&	&	&	&	I_{32}&	&	&	\\
\hline
	&	&	&	&	&	&	&	&	&	&	&	&	&	I_{32}&	&	\\
	&	&	&	&	&	&	&	&	&	&	&	R^{13}&	&	&	I_{32}&	\\
	&	&	&	&	&	&	&	&	&	&	&	&	&	&	&	I_{32}\\
\hline
I_{32}&	&	R^{10}&	&	&	&	&	&	&	&	&	&	&	&	&	\\
\end{array}\right)
\]
\caption{Transition matrix of a word oriented LFSR}\label{lfsr-word-mat}
\end{figure*}

\end{document}